\documentclass[superscriptaddress,aps,secnumarabic]{revtex4}

\usepackage[latin2]{inputenc}
\usepackage{amssymb,amsmath}
\usepackage{graphicx,float}
\input{psfig.sty}

\begin{document}

\title{ \Large \bf Transport on Complex Networks: Flow, Jamming and Optimization}

\author{Bosiljka Tadi\'c}
\email{bosiljka.tadic@ijs.si}
\affiliation{Department  for Theoretical Physics, Jo\v{z}ef Stefan Institute, 
P.O. Box 3000, 1001  Ljubljana, Slovenia}
\author{G. J. Rodgers}
\affiliation{Department of Mathematical Sciences, Brunel University, Uxbridge, Middlesex UB8 3PH; UK}
\author{Stefan Thurner}
\email{thurner@univie.ac.at}
\affiliation{Complex Systems Research Group, HNO, Medical University of Vienna, W\"ahringer G\"urtel 18-20, A-1090 Vienna, Austria}

\begin{abstract}
\noindent
Many transport processes on networks depend crucially on the underlying network geometry, 
although the exact relationship between the structure of the network and the 
properties of transport processes remain elusive.
In this paper we address this question by using numerical models in which both 
structure and dynamics are controlled systematically. We consider the traffic of information packets 
that include driving, searching and queuing.  We present the results of extensive 
simulations on two classes of networks; a correlated 
cyclic scale-free network and an uncorrelated homogeneous weakly clustered network. 
By measuring different dynamical variables in the free flow regime we show how the 
global statistical properties of the transport are related to the temporal fluctuations 
at individual nodes (the traffic noise) and the links (the traffic flow). We then demonstrate 
that these two network classes  appear as representative 
topologies for optimal traffic flow in the regimes of low density and high density traffic, 
respectively. We also determine statistical indicators of the pre-jamming regime on 
different network geometries and discuss the role of queuing and dynamical betweenness 
for the traffic congestion. The transition to the jammed traffic regime at a critical 
posting rate on different network topologies is studied as a phase transition with an 
appropriate order parameter. We also address several open theoretical problems related 
to the network dynamics. 
\end{abstract}

\maketitle

\section{Introduction \label{sect_intro}}

\subsection{Motivation} 
Networks form part of the basic model for virtually 
every known co-operative phenomenon, and the study of processes on networks 
is fundamental to a large part of the physical, biological, social, 
economic and engineering sciences \cite{blmch-cnsd-06}.  
Most technological, biological, economical or social 
networks support a number of transport processes, such as the traffic of 
information packets 
\cite{tr-ptsfn-02,tt-isdsn-04,ttr-tcntugspmdf-04,adg-cnhb-01,gadg-dpmcn-02,sv-itptmit-01,vs-soctpcn-02,os-ptcntm-98,mpvv-clcisfn-03,rosevall05,rosevall03,valverde04,wwyxz-tdblrpsfn-06}, 
signals, molecules \cite{agrawal02}, finance and wealth \cite{thurner03,thurner04,coelho05}, 
rumours \cite{moreno04} or diseases \cite{newman02,bpv-escndc-03}. One can model 
these processes in many different ways, from simple interacting random walkers 
\cite{t-arwcwg-01,noh04,eisler05}, 
to interacting random walks with random traps \cite{gallos04} on the network that serve as 
destinations, to richer models in which the traps are walker specific. Some of 
the models we discuss here allow an even wider range of behaviour; each walker 
has a destination which is known and fixed at the start, potentially allowing 
both local and global search and navigation strategies to be introduced to 
improve the efficiency of the transport.

In recent years, studies of different networks have revealed that
the dynamic processes crucially depend on the topology of the underlying 
network structure \cite{ttr-tcntugspmdf-04,gadg-dpmcn-02}. 
However, the precise interdependencies between the functional properties and the 
relevant structural parameters, which define network classes, remain largely unknown. 
In  naturally evolved networks, such as the cytosceleton in cells, 
which serve as a backbone of molecular transport within the cell, efficient 
functioning emerges through an optimisation of function via the evolutionary 
adjustment of the structure.  This structure--function optimization  is an 
imperative in artificial networks because of the need for increased efficiency
\cite{lm-ebswn-01,jafj-astifon-05}, 
low risk of information loss, low levels of congestion \cite{gdvca-ontlsc-02,mpvv-clcisfn-03} and low risk of   
critical malfunction \cite{holme03,echenique05,ashton05}. 
Therefore, understanding a network's functional properties 
and their dependence on a {\it reduced} set of its structural parameters 
is of paramount importance.
This understanding is being developed through the systematic collection of 
empirical data, and through the introduction, simulation and 
solution of new algorithms that seek to improve one or more aspects of the 
functional performance of the network. This work is providing the basis for a 
deeper theoretical understanding of transport on networks, that allows the 
identification of universality classes of both network topologies and transport 
processes.

\subsection{Methodology}
This paper reviews some of the work that  has been  carried out to progress this agenda. 
In our approach both the network structure and the dynamics on the network are 
controlled systematically within a numerical model of information traffic. 
In this model the posting, navigation and queuing of packets are parameterized 
by a set of control parameters. Then the properties of the traffic are determined 
for different underlying network structures. Two approaches are available to 
optimise traffic. When the network structure is fixed on the time scale of 
the transport, one can find sets of optimal paths between nodes and specify 
the traffic along these paths. Alternatively, when the network structure is 
allowed to evolve on the same time scale as the transport, an optimal network 
structure is found that depends on the traffic conditions.

\subsection{Properties of Information Traffic} 

The traffic of information packets is defined via several properties and parameters. 
Here we summarize these properties, which are later realized in the simulations. 

\begin{itemize}
\item {\it Creation \& Assignment}. 
An information packet is created, at some rate, at a node $i$ and is assigned 
the address of another node $j$ where it should be delivered; When a packet 
arrives at its destination it is removed from the network; 

\item {\it Traffic Queues}. 
Packets form queues when two or more packets are at same node at 
the same time; The queue length at node $i$ at time $t$ is denoted by $Q_i(t)$; 

\item {\it Queuing Discipline}. 
This determines the order in which packets leave a queue; In this work we use 
the last-in-first-out (LIFO) queuing discipline for traffic  simulations;  

\item {\it Waiting Time $t_w$}. 
Is the time that a packet spends in a particular queue waiting to leave;

\item  {\it Travel Time $T$}. 
Is the total time that a packet spends in the network from its creation 
to its delivery at its destination. This time is equal to the summation of the 
path length and the waiting time at each queue along the  path; The travel 
time of a packet is related to its {\it travel costs}, sometimes called the 
{\it search costs}.

\item {\it Navigation \& Search Depth}. 
Information about a packet's destination is required if one wishes to reduce 
its travel time by using a navigation process; Global navigation is a costly procedure 
in which a shortest (or the best) path for each packet is determined;  A much less 
costly alternative is to use {\it local search} algorithms in which each packet explores the 
neighbourhood of its current node for its destination address or for an optimum 
direction;  A search depth of $d=1$ corresponds to the nearest-neighbourhood of 
a node and $d=2$ corresponds to the next-nearest-neighbourhood a node, sometimes 
called $nnn$-search; A search depth of $d=0$ corresponds to random processing;
In this paper we employ $d=2$, $nnn$-search;

\item  {\it Dynamic Load}. 
The number of packets transported by a node $i$, or along a link $ij$, 
at time $t$ defines the dynamic load of the node $h_i(t)$ 
and the link $f_{ij}(t)$; The total number of packets in the network 
$N_p(t)$ is the network load, obviously $N_p(t) =\sum _iQ_i(t)$, the 
sum of all queue lengths at time $t$;

\item  {\it Traffic Noise}. 
The time signals of temporally fluctuating variables such as 
$h_i(t)$, $N_p(t)$, the number of simultaneously active nodes $n(t)$, etc. 
are often called traffic noise signals. In particular, the set of signals
 $\{h_i(t)\}$ recorded at each of $i=1,2, \cdots , N$ nodes simultaneously 
form the basis for {\it multi-channel noise analysis};  Analogously,
{\it traffic flow} is the same quantity defined for links rather than nodes;

\item {\it Free Flow Regime}. Refers to the free (uncongested) flow of 
packets, which is compatible with {\it stationarity} of the traffic noise 
time-series;

 \item {\it Congested Regime}. 
 Corresponds to a partial or complete jamming in networks when packets can get stuck 
 for an indefinite time; The network load  increases steadily with time, 
 making the time-series non-stationary.

\end{itemize}

\subsection{Concepts}

We investigate our numerical model for information packet transport on two types of networks; a 
clustered scale-free network and a homogeneous network, which are, respectively, 
representatives 
of the {\it causal} and {\it homogeneous} \cite{bjk-pgun-04} network classes.  
The topological properties of these networks that are relevant to 
transport process are discussed in detail in Section \ref{sect_flow_simulations}, 
along with the transport rules. 
We then present the results of simulations of traffic on 
these networks. These results consist of statistical properties of traffic  
collected on the global network level (such as probability distributions of packet travel times
and waiting time, etc.), and  local (individual nodes and links) activity during transport, 
sometimes called multi-channel noise and flow analysis. 
In Section\ \ref{sect_jamming} the statistical signatures of traffic jamming 
are discussed numerically for both 
network types. In addition, we consider the transition to the congested 
traffic phase, where travel and waiting times
tend to diverge, as a dynamical phase transition. 
Section\ \ref{sect_optimization} discusses two optimization procedures, one with fixed network geometry, 
as above, and the other which involves the network restructuring in order to minimize the 
packet
travel times. We show that the emergent optimized structures 
in low and high density traffic are statistically similar to the structures discussed in 
Section \ref{sect_flow_simulations}.
In Section\ \ref{sect_openproblems} we give a summary of the results and some 
open theoretical problems posed by the numerical simulations and by empirical measurements 
in real communication networks.

\section{ Stationary Traffic Flow \label{sect_flow_simulations}}

In the first part of this section we introduce two networks---a grown 
scale-free
correlated network, which we call the Webgraph, and a homogeneous network with a stretched-exponential
degree distribution, which we call Statnet. We briefly summarise the structural 
properties of these networks that are relevant to their transport processes.
 We then introduce the model of 
packet transport and perform simulations on these networks with $N=1000$ nodes.
The results we obtain for the statistical properties of the transport in the {\it free 
flow} regime are presented. 

\subsection{Structural Properties of the Networks}

We consider two networks assembled  by {\it preferential attachment}
\cite{dm-en-03} and in which {\it rewiring} \cite{t-ddgwww-01} occurs. 

\begin{figure}[thb]
\begin{center}
\begin{tabular}{cc} 
{\psfig{file=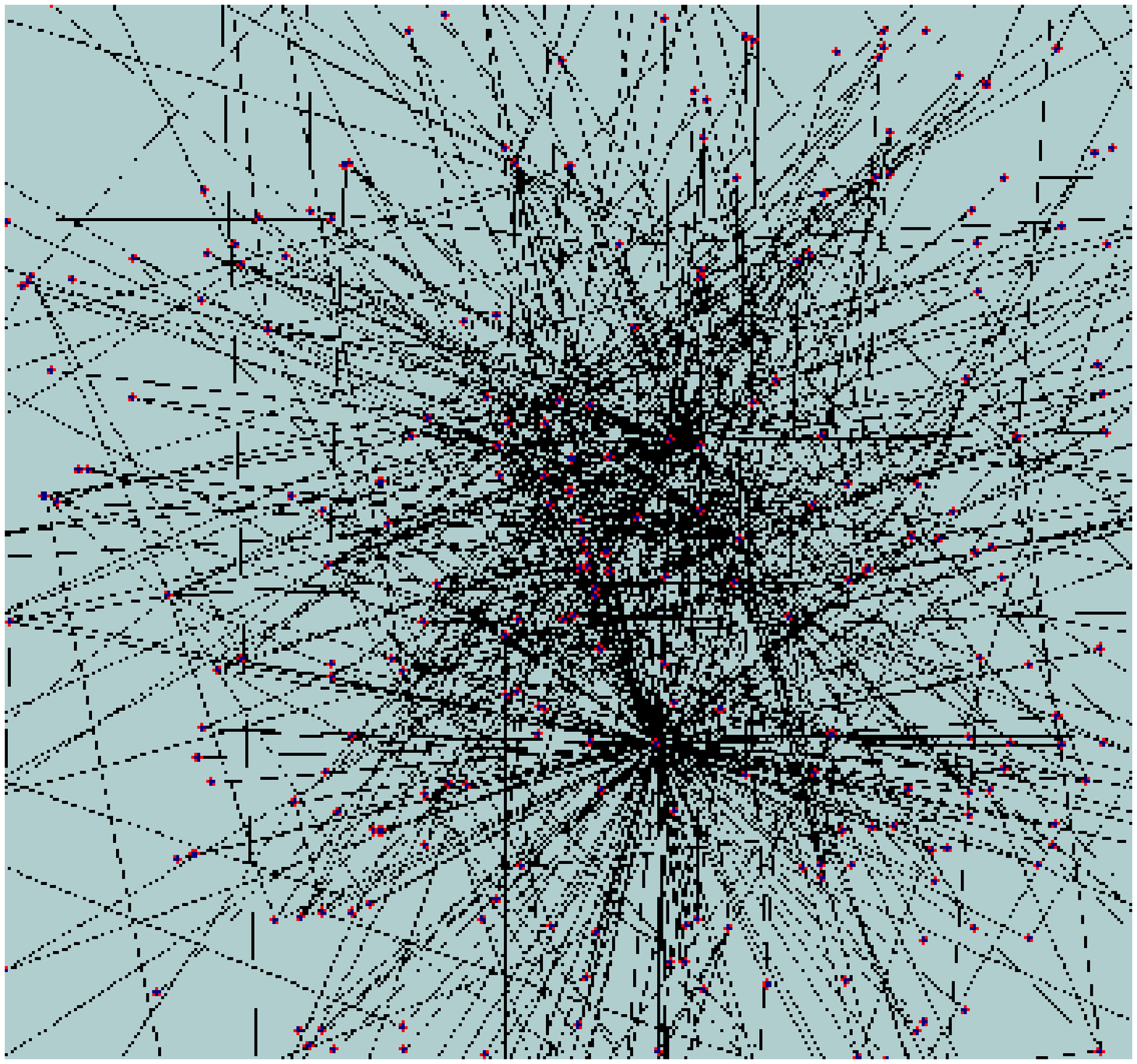,height=6.4cm,width=6.4cm}}&
{\psfig{file=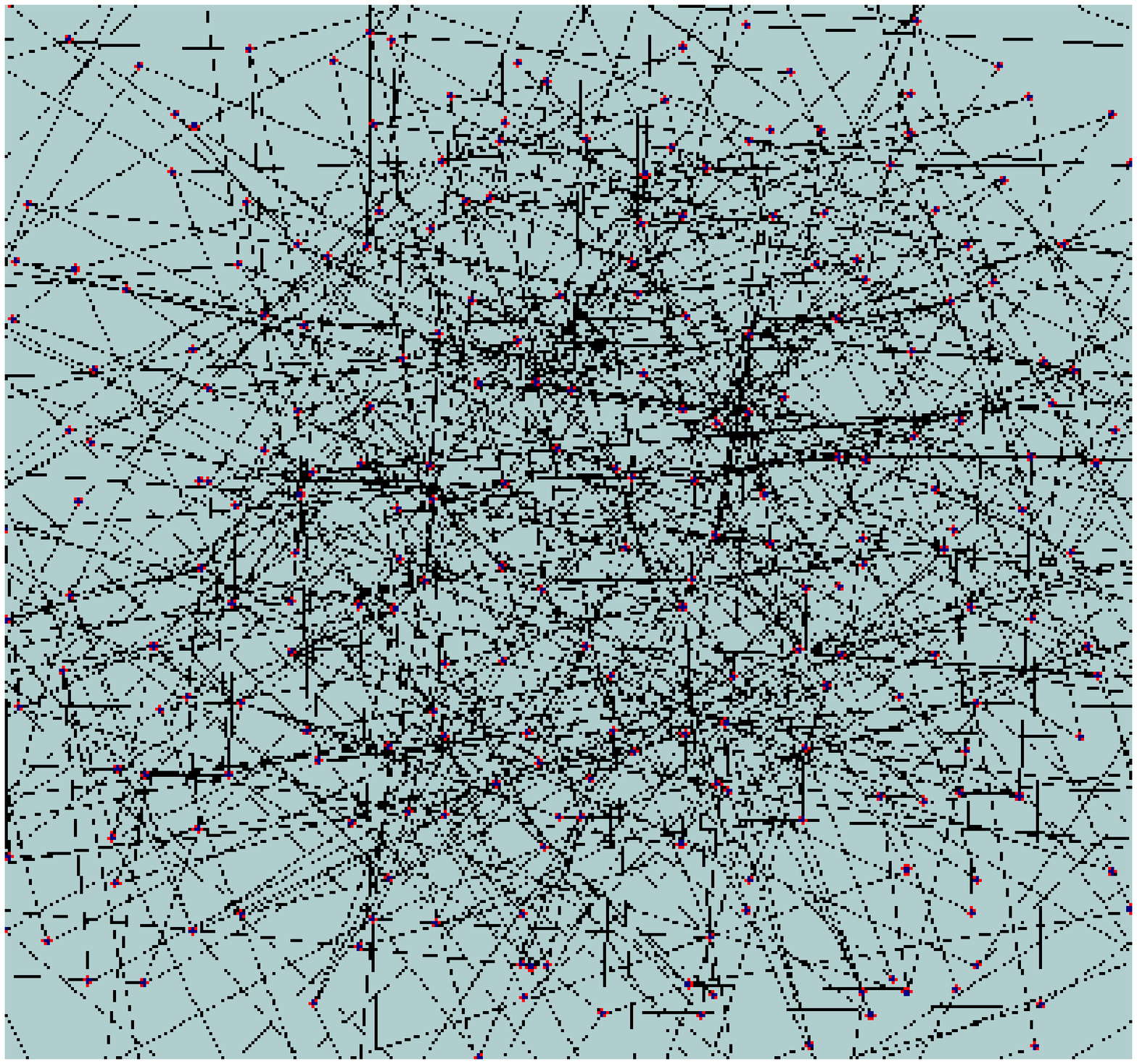,height=6.4cm,width=6.4cm}}\\
{\Large $(a)$} &{\Large $(b)$}
\end{tabular}
\end{center}
\caption{ (a) The cyclic scale-free Webgraph and (b) homogeneous Statnet.}
\label{figgraph}
\end{figure}

The Webgraph is grown by {\it sequentially adding nodes}  
from $i=1,2, \cdots N$, with linking rules that involve 
preferential attachment and preferential rewiring 
according to the time dependent probabilities $p_{in}$ and $p_{out}$. 
They are 
applied in the subset of {\it pre-existing} nodes at each growth step $i$.
The linking probabilities depend on the current number of incoming $q_{in}$ and 
outgoing $q_{out}$ links at a node \cite{t-ddgwww-01}
\begin{equation}
p_{in}(k,i) = \frac{\alpha + q_{in}(k,i)/M}{(1+\alpha )i}  \quad \quad
p_{out}(n,i) =\frac{\alpha + q_{out}(n,i)/M}{(1+\alpha )i}  \quad .
\label{pinpout}
\end{equation}
Linking to a new node occurs with probability ${\tilde{\alpha}}$ and rewiring or 
adding a link from a previously 
existing node occurs with probability $1-{\tilde{\alpha}}$. These competing 
processes lead to an emergent structure \cite{t-ddgwww-01} which is different 
to the structure obtained by 
common  preferential attachment, \cite{ab-smcn-02,dm-en-03}. 
The parameters of the model  ${\tilde{\alpha}}$, $\alpha $, and $M$ are responsible 
for the graph's flexibility,
connectivity profile and clustering. In particular, when ${\tilde{\alpha}} =1$ the 
graph is rigid (rewiring can not occure)  and scale-free, i.e., for $M=1$ a scale-free tree and 
when $M>1$ a weakly clustered uncorrelated scale-free graph emerges as known in 
Refs.\ \cite{ab-smcn-02,dm-en-03}. 
However, when ${\tilde{\alpha}} < 1$, these rules lead to power-law 
distributions of both in-degree and out-degree \cite{t-ddgwww-01}, a large 
clustering coefficient and link correlations \cite{t-mtipgct-03}, even for 
$M=1$.
When  ${\tilde{\alpha}} = \alpha = 1/4$ the structure is very similar to the 
world wide web \cite{t-ddgwww-01}. This is why we call this network the Webgraph. 
An example of the emergent structure is shown in  Fig.\ \ref{figgraph}a. 

To grow the Statnet we apply the same rules with the probabilities in 
Eq.\ (\ref{pinpout}), however, with the fixed number of nodes $i=N$, among 
which {\it links are added sequentially}.  Multiple linking between the 
same pair of nodes is not allowed. The Statnet emergent structure, with 
$L=N= 1000$, is also shown in Fig.\ \ref{figgraph}b.

A detailed quantitative analysis of the structure reveals that both 
incoming and outgoing links behave the same statistically, and exhibit a 
stretched-exponential degree distribution. In addition, the clustering in 
this graph is small compared to the Webgraph and link correlations 
are entirely absent.  A comparison of the structural properties 
of both networks is given below.

\begin{figure}[htb]
\begin{center}
\begin{tabular}{cc} 
{\psfig{file=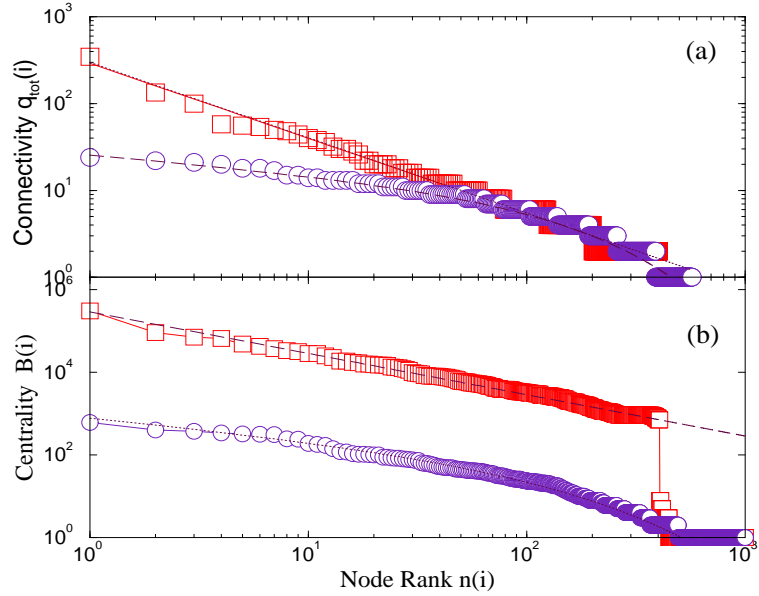,height=8cm,width=10cm}}\\
\end{tabular}
\end{center}
\caption{(a) Node ranking of the total connectivity 
$q_{tot}=q_{in}+q_{out}$ and  (b) node ranking of the topological centrality $B(i)$ in the Webgraph ($\square $) and Statnet  ($\bigcirc$) structures. The fits are explained in the text.}
\label{fig_conn-centrality}
\end{figure} 
\noindent
{\it Connectivity:}  The emergent connectivity of nodes in the evolving 
Webgraph can be obtained analytically using linking probabilities in
Eqs.\ (\ref{pinpout}). Both for in-coming and out-going connectivity we have a power-law profile according to \cite{tp-vdgcn-02}
\begin{equation}
q_{\kappa}(s,N) = A_\kappa \left[\left(\frac{N}{s}\right)^{\gamma _\kappa} -B_\kappa\right] \ ;
\label{qkappas}
\end{equation}
after $N$ nodes are added, $s$ being the addition time.
Here $\kappa $ indicates 'in', 'out', or 'total' links, for which 
different exponents are found \cite{tp-vdgcn-02,t-tfsoplwww-02}. For the purposes of this work, we consider that the transport along a link in both directions is symmetrical. Therefore, the total node connectivity $q_{tot} \equiv q_{in} + q_{out}$ is relevant.  In Fig.\ \ref{fig_conn-centrality}a we show the connectivity profile of all nodes in one Webgraph sample (shown in Fig.\ \ref{figgraph}a). It exhibits a power-law decay with rank $n\equiv n(i)$  of a node $i$ according to 
\begin{equation} 
X(n) \sim n^{-\gamma_{tot} } \ ;
\label{pl_n}
\end{equation}
with $\gamma_{tot} = 0.867$. This indicates a power-law connectivity distribution with the exponent $\tau _{tot} = 1/\gamma _{tot} +1 = 2.153$. Similarly, node ranking according to the total connectivity in the Statnet, also shown in Fig.\ \ref{fig_conn-centrality}a, is well fitted by a stretched-exponential curve 
\begin{equation}
X(n) = An^{-\tau }\exp{[-(n/n_0)^{\sigma}]} \ ;
\label{stretch-exp}
\end{equation}
with $\tau \approx  0.22$,  $\sigma  \approx0.78$, and $n_0 \approx 200$.
The emergent probability distribution also obeys this law, Eq. (\ref{stretch-exp}).

\noindent
{\it Centrality:} Further quantitative differences between these two networks are found in the node ranking using the topological centrality measure \cite{n-scnspwnc-01,zlw-cscn-05,Latora04}. 
In Fig.\ \ref{fig_conn-centrality}b the ranked profile of topological betweenness of nodes are shown. For the Webgraph the profile is a power-law Eq.\ \ref{pl_n} with the exponent $\gamma _B =1$, indicating the distribution of node betweenness as $P(B) \sim B^{-2}$.
In the case of Statnet the profile is again closer to a stretched-exponential form Eq. (\ref{stretch-exp}). 
In Fig.\ \ref{fig_conn-centrality}b one can also identify the 
nodes belonging to the giant cluster, whose size is 403 nodes in the Webgraph, and 571 in the Statnet (jump). 

\begin{figure}[htb]
\begin{center}
\begin{tabular}{cc} 
{\psfig{file=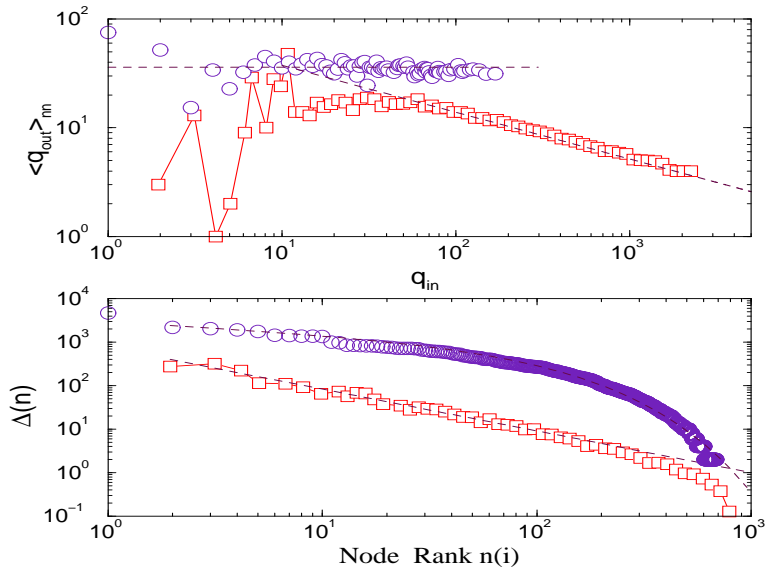,height=3in,width=4.0in}}
\end{tabular}
\end{center}
\caption{Link-correlations (top) and ranked clustering profile (bottom) for
the Webgraph ($\square $) with $L/N=1$ and Statnet ($\bigcirc$) with $L/N=20$  corresponding to the 
same average clustering coefficient, $cc=0.36$.}
\label{fig_ck_clustering}
\end{figure} 

\noindent
{\it Clustering:}  Due to the rewiring, some nodes and/or smaller clusters remain disconnected from the giant component, whereas other nodes
gain large connectivity and clustering. When the average number of links per node is $M\equiv L/N=1$ the average clustering coefficient in the Webgraph appears to be $cc=0.3601$. In contrast Statnet, with same average connectivity $M=1$, 
has a much lower average clustering $cc=0.0075$, which increases with $M$. The clustering profile is 
inhomogeneous in both networks. It obeys a power-law, shown in 
Fig.\ \ref{fig_ck_clustering}a with an exponent close to $\gamma =0.85$ in  the Webgraph, where
 most of the elementary triangles are attached to the main hubs. 
In the case of Statnet a stretched-exponential profile is found. 

\noindent
{\it Correlations:}
Another interesting feature which might influence the transport processes 
with search is found in the link correlations  of the Webgraph, which are entirely absent in Statnet. 
The computed correlation property $\Delta_i$, denoting the number of elementary 
triangles attached to a node $i$, is shown
in Fig.\ \ref{fig_ck_clustering}b for both graphs after ordering. The power-law 
decay with the exponent $\gamma _c=0.42$ in the Webgraph indicates 
{\it disassortative} \cite{n-mpn-03} link correlations.

\subsection{Implementation of Constant-Density Traffic}

We model the traffic of information packets on a network as a 
{\it guided random walk} between specified pairs of nodes
\cite{tt-isdsn-04,ttr-tcntugspmdf-04,t-mtipgct-03}---the origin and 
destination (delivery address) of a packet. 
Once created, packets navigate through the network using a local 
$nnn$-search rule \cite{tt-isdsn-04,tt-statsfn-05,t-mtipgct-03}, in which 
the next-nearest-neighbourhood is searched for the destination node. 
If a node finds that a packet's destination node is in its nearest 
neighbourhood, 
it moves the packet  directly to its destination. 
If the destination node is in the next-nearest-neighbourhood, but not the 
nearest-neighbourhood, the packet moves in the correct direction. 
If the destination node isn't found in the next-nearest-neighbourhood, 
the packet moves to a randomly chosen neighbouring node. 
Packets are removed when they arrive at
their destinations. In this section we model traffic with a
{\it fixed number of moving packets}, which is set at the start of the 
simulation, so that $N_p(t)=\rho$, fixed for all $t$. Packets that arrive at 
their destinations and are removed 
from the network are replaced in the next time step
by the creation of the same number of new packets at randomly chosen nodes. 
Of course, these packets are given new destinations. 
When $\rho =1$, 
this corresponds to the sequential random walk problem. 
At density $\rho > 1$, packets interact by forming queues at 
nodes along their paths. We assume finite maximum queue lengths of $H=1000$
at all nodes, 
and we employ a LIFO (last-in-first-out) queuing discipline. 

Constant-density traffic is interesting because the limit of non-interacting 
packets $\rho =1$ is mathematically correctly implemented. 
This enables quantitative analysis of the structure--dynamics interdependences 
without additional dynamic effects \cite{tt-statsfn-05}. 
Also, at finite traffic density $\rho > 1$, the system is driven {\it self-consistently}, without external
forcing. This situation is suitable for the analysis of time-series and
noise fluctuations (see later), which are sensitive to driving modes.
Other driving conditions, e.g., constant posting rate $R$ 
\cite{tt-isdsn-04,ttr-tcntugspmdf-04,adg-cnhb-01,acdgv-sccn-03}, 
will be considered in Section\ \ref{sect_jamming} in connection 
with traffic jamming. When the traffic is driven by creating at 
each time step a number of packets which is larger than network's 
output rate, the network experiences congestion \cite{gadg-dpmcn-02}. 
Driving at constant rate $R$ is appropriate for a quantitative study of
the congestion problem (see Section\ \ref{sect_jamming}).

For the numerical implementation of the transport, we first generate the 
network and its  adjacency matrix is stored and remains fixed throughout the transport
process. If the graph is disconnected, as is the case with both Webgraph and Statnet, we take 
care that the creation and destination nodes are within the same cluster. 
We study networks consisting of $N=1000$ nodes. 
Starting with $\rho =100$ packets, which are  created at random 
positions, the network is updated in parallel. At each time step a node with a packet on it tries  
to move the top packet in its queue towards that packet's destination node. 
A packet is moved to one of node's neighbours, and joins 
the top of its queue. If that neighbour is the packet's destination node, the packet 
is considered as delivered and disappears from the network.
During the transport process, for each packet we keep track of 
its destination,  current position, and position in the current queue.
In addition, for a subset of labeled, $N_M=2000$, packets we keep track of the time that 
they spend in each  queue before they arrive at their destination.
We compute the statistical properties of the transport from the data gathered 
from the labeled packets once all of them have arrived at their destinations.

\subsection{Global Transport Characteristics}
The statistical properties of traffic depend on both the network structure
and traffic conditions. For a fixed navigation protocol ($nnn$-search in this case), 
the efficiency of transport depends on the
 structural characteristics of the underlying network and on  
the overall traffic density.  In particular, the travel time of packets
is determined  by the length of the path selected between the origin and 
destination node, and the waiting times at nodes along that path. 
That is, for a packet traveling along a path of length $k$, the travel time is given by
\begin{equation}
T_k = \sum _{i=1}^k t_w(i) \ ,
\label{def_Tk}
\end{equation}
where $t_w(i)$ is a packet's waiting time on node $i$ on that path. It should be stressed that both path lengths and set of waiting times $\{t_w\}$ 
which contribute to the travel time $T$ are outcomes of a stochastic process.
Then the  functional central-limit theorem \cite{whitt01}  
applies to the distribution of travel times $P(T)$, computed along all paths on the network \cite{tt-isdsn-04}. 
The fact that on structured networks the waiting times along a particular path depend on the identity 
of nodes along that path adds to the complexity of the problem 
\cite{tt-isdsn-04}.

\begin{figure}[htb]
\begin{center}
\begin{tabular}{c} 
{\psfig{file=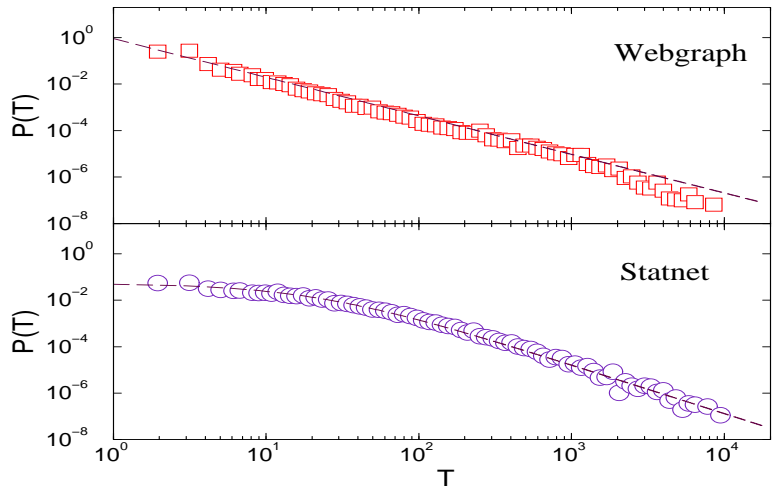,height=2.6in,width=4.0in}}
\end{tabular}
\end{center}
\caption{Distribution of travel times  in the Webgraph and the Statnet for  a density 
 $\rho =100$.}
\label{fig_ptt}
\end{figure}
In the limit of non-interacting packets $\rho =1$, the waiting times are
$t_w=1$ for all nodes, thus the entire travel time is 
determined by the geometry. The suitability of the navigation algorithm for a given topology 
can then be measured by the deviation of the actual path of the packet 
from the {\it shortest path } between the pair of nodes.
In Fig.\ \ref{fig_ptt} we show how the distributions of the travel times for
the Webgraph and Statnet geometries and $\rho =100$ packets. 
For $nnn-$search these distributions exhibit power-law tails with different
 exponents on the two networks. However, the largest difference appears 
for short travel times,  where packets follow more closely the topological 
shortest path on the underlying network. In this respect the node connectivity and centrality play an important role 
 (see Fig.\ \ref{fig_conn-centrality}).
The deviations from the shortest paths are experienced by packets which, i.e.  local search being
ineffective, appear to perform a random walk in
 parts of the graph far away from their destinations. 
The distributions of travel times at large density $\rho =100$ appear to 
be well fitted with a power-law 
\begin{equation}
P(T) = AT^{-\tau _T} \ ;
\label{ptt_pl}
\end{equation}
with  $\tau _T\approx 1.5$ and an exponential cut-off (due to the finite network size) for packets on the Webgraph, and a $q-$exponential
distribution 
\begin{equation}
P(T) = B_q\left[1 + (1-q)T/T_0\right]^{1/(1-q)} \ ;
\label{PT_qexp}
\end{equation}
 with $q=1.47$,
for the case of transport on the Statnet (cf. Fig.\ \ref{fig_ptt}).

\begin{figure}[htb]
\begin{center}
\begin{tabular}{c} 
{\psfig{file=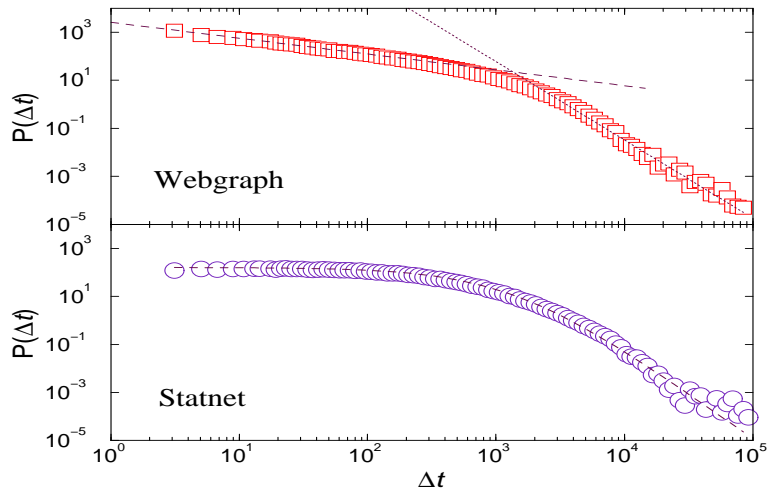,height=2.6in,width=4.0in}}
\end{tabular}
\end{center}
\caption{Distribution of return  times of packets in Webgraph  and 
Statnet  for  a density $\rho =1$.}
\label{fig_prett}
\end{figure}

In processes in which diffusion dominates, the topology of actual paths gives 
another view of the process. 
In particular, on structured networks, apart from the inhomogeneous connectivity, 
the number of short and long cycles may effect the  diffusion of packets.
The network  profile of the short cycles (triangles) in the Webgraph and Statnet 
are shown in Fig.\ \ref{fig_ck_clustering}.
The packets that perform a random diffusion, i.e. when the local search is 
for them ineffective, may return to a node, which had already been
visited in the past. The distribution of the return-time intervals, 
$\Delta t_i$, 
for each node $i=1,2, \cdots  N$ on the network, is 
given in Fig.\ \ref{fig_prett} for both network types. 
Broad distributions with characteristic power-law tails,  suggest correlated events.
 Again the main difference 
between the two network types appears in the small  
return times of packets. A power-law with small slope $\tau _\Delta =0.66$
 for $\Delta t \lesssim 800$
 can be related with an 
uneven population of short cycles and the active role of individual nodes 
in the case of Webgraph. 
In the case of Statnet the distribution can be fitted to the general expression in
Eq.\ (\ref{PT_qexp}), with $q=1.26$. 
On the Webgraph, the probability of long return times falls off as $P(\Delta t) \sim (\Delta t)^{-3.26}$. 
When the packet density is finite, the return time of node activity
(where generally a different packet is involved) is different on the two networks. The distribution is shifted towards shorter values in the case of 
Webgraph. On Statnet the form of the distribution changes.

\begin{figure}[htb]
\begin{center}
\begin{tabular}{c} 
{\psfig{file=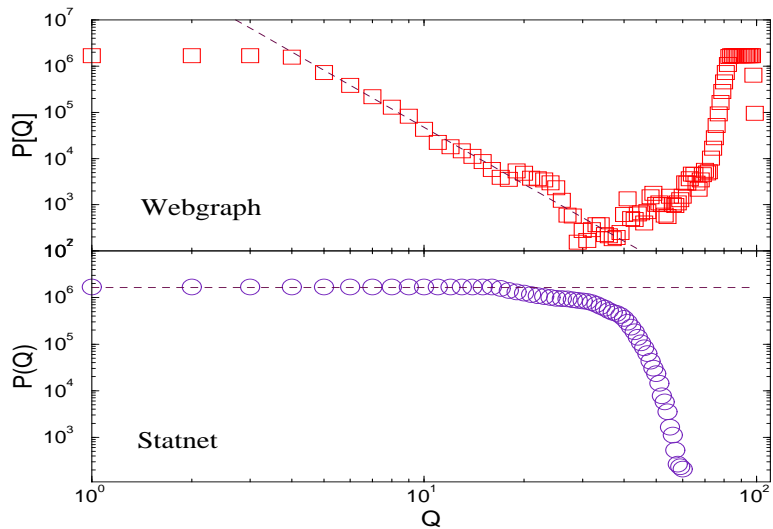,height=2.8in,width=4.0in}}
\end{tabular}
\end{center}
\caption{Distribution of queue-lengths in  Webgraph and  Statnet for a  packet
density  $\rho =100$.}
\label{fig_pqueue}
\end{figure}
For large packet density $\rho >> 1$, motion of a packet will be 
affected by other packets moving through the  same node, as mentioned above. 
Queues 
of packets then occur and a queuing discipline sets the order of 
processing (LIFO in this particular case). Apart from the traffic 
density, the neighbourhood of the node determines the length 
of the queue at that node. In particular, on inhomogeneous networks, 
hubs appear to receive  more packets compared to other nodes, due to 
their large connectivity. Since in the algorithm one packet is 
processed per time step, other packets remain in the queue to be 
processed later (when no new packet has been received). 
The distribution of queue lengths therefore reflects the 
network structure in a particular way. 
A snapshot of queue-lengths $Q$, for traffic density $\rho = 100$ in the 
two network structures leads to the distributions shown in 
Fig.\ \ref{fig_pqueue}. 

 In the homogeneous Statnet most of the nodes are processing a similar number of packets,
which leads to a flat distribution of queues and a cut-off indicating that queues longer than 
$Q\approx 40$ rarely occur. On the other hand, a large queue of 
$Q=80$ to $90$, 
packets can be  found on the hubs on the inhomogeneous
 Webgraph  with high probability. On the rest of the nodes the queues are distributed 
 with a power-law distribution, apart from very small queues at boundary  nodes.
\begin{figure}[htb]
\begin{center}
\begin{tabular}{c} 
{\psfig{file=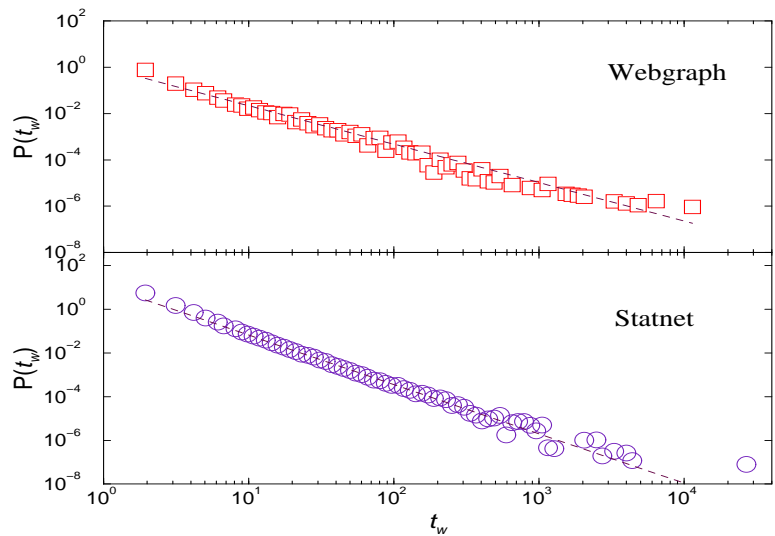,height=2.8in,width=4.0in}}
\end{tabular}
\end{center}
\caption{Distribution of waiting times in the Webgraph and the Statnet for  a density 
 $\rho =100$.}
\label{fig_ptw}
\end{figure}
 
The queuing times of packets extend their travel times, thus reducing the overall traffic 
efficiency \cite{ttr-tcntugspmdf-04}. Note, that in the current 
implementation, with a constant number of moving packets, jamming cannot occur as 
long as the traffic density $\rho < H$, where $H$ is the 
maximum allowed queue length. However, due to long waiting times in queues, 
the travel times of packets given by Eq.\ (\ref{def_Tk}), can become very long, 
given by a power-law distribution, as shown in Fig.\ \ref{fig_ptt} (see also
 \cite{tt-isdsn-04,tt-statsfn-05} for different  types of graphs and search
algorithms).  
The distributions of waiting times of packets on two network geometries are given 
in Fig.\ \ref{fig_ptw}, for the traffic density $\rho =100$. They can be 
described as power-law distributions,
\begin{equation}
P(t_w) \sim At_w^{-\tau_w} \ ;
\label{ptw}
\end{equation}
 with different  exponents, closely related 
to the tails of the travel-time distributions in Fig.\ \ref{fig_ptt}. 
Specifically, in the case of Statnet the numerical value of the exponent  is $\tau_w \geq 2$,
suggesting finite average waiting times, and thus finite travel times 
for packets on this network structure. 
In contrast, the average waiting time on 
the Webgraph for this traffic density is not bounded in a mathematical sense, 
having $\tau_w < 2$. This implies a systematic increase of the travel times 
of packets on a large network with this structure (with large measurement 
time). The distribution of waiting times and travel times of individual 
packets provide a quantitative measure that supports the conclusion that 
the Webgraph class of networks, although much more efficient compared to 
other scale-free network types \cite{tt-isdsn-04,ttr-tcntugspmdf-04}, 
is the less efficient of the two networks  at high constant density of packets. 
More homogeneous structures, such as the Statnet, appear to perform 
better under these traffic conditions. It should be stressed that the waiting times as 
well as travel times are traffic properties measured for individual packets,
which will depend on the type of queuing discipline used.

\subsection{Noise and Flow on Networks \label{sect_noise}}

\begin{figure}[htb]
\begin{center}
\begin{tabular}{c} 
{\psfig{file=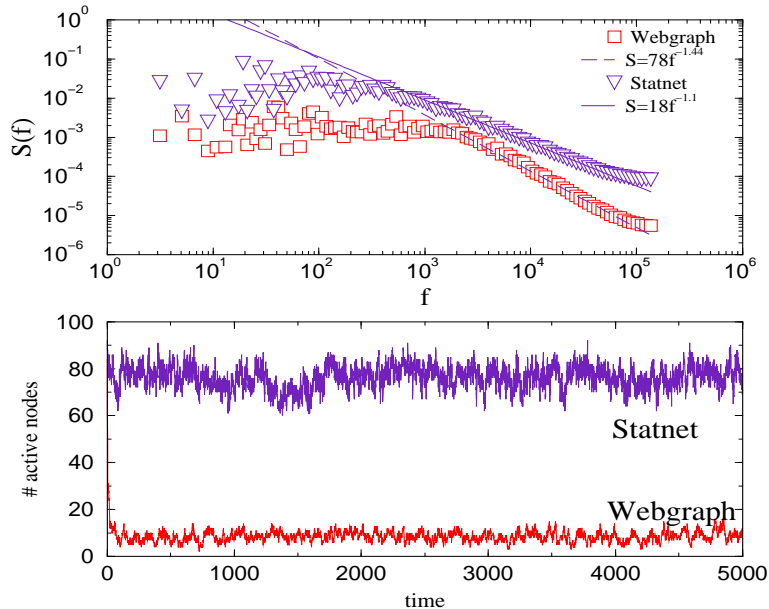,height=3.2in,width=4.0in}}
\end{tabular}
\end{center}
\caption{Temporal fluctuations of the number of active nodes in Webgraph and 
Statnet  
(lower panel) and their power spectra (top panel) for a density of $\rho = 100$ 
packets (from [Tadic and Thurner, 2006]).}
\label{fig_psact}
\end{figure}

The observed queue lengths are compatible with the the temporal 
properties of node activity on the two networks, shown in Fig.\ \ref{fig_psact}. 
While queues at important nodes in the inhomogeneous Webgraph  are long, 
the number of nodes that are simultaneously active is small, fluctuating 
about an average value $n_a\approx  8$.
Compared to the more homogeneous Statnet for the same traffic density, on 
average $n_a \approx 80$ nodes are processing a packet simultaneously, 
leading to short queues at all nodes. 
Further quantitative analysis 
of the time-series reveals the differences in the packet processing of 
the two classes of networks. In particular, long-range correlations 
(anti-persistence) in the number of active nodes develops on both networks. 
However, in the conditions of constant packet density the fluctuations on 
the Statnet  appear to be more correlated compared to the Webgraph. 
The power spectrum exhibits an asymptotic power-law behavior 
\begin{equation} 
S(f) \sim f^{-\phi} \quad , 
\label{def_ps}
\end{equation}
shown in Fig.\ \ref{fig_psact} (top panel),
with $\phi =1.1$ for Statnet and $\phi = 1.4$ for the Webgraph. 
Therefore, an 
increased traffic density leads to stronger correlations in node 
activity on the more homogeneous Statnet \cite{tt-tnmfstgsn-06}. 

\begin{figure}[htb]
\begin{center}
\begin{tabular}{c} 
{\psfig{file=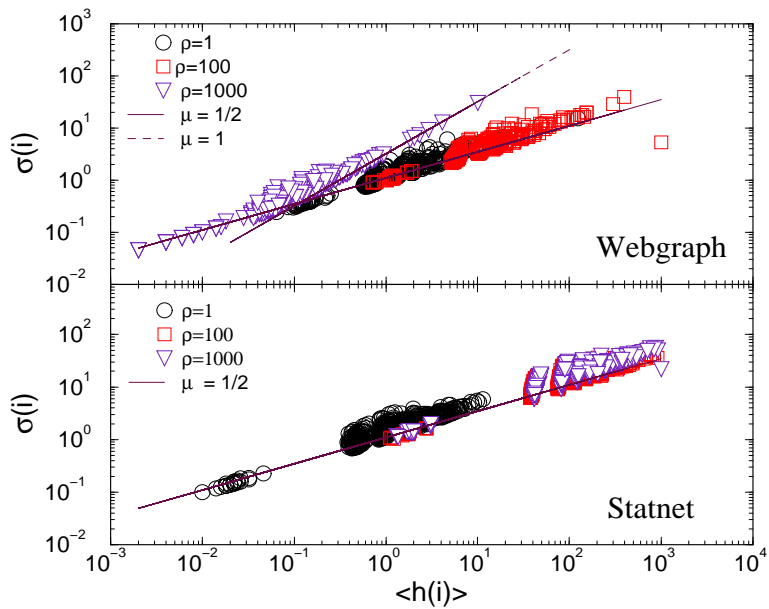,height=3.2in,width=4.0in}}
\end{tabular}
\end{center}
\caption{Universal noise fluctuations for traffic with constant-density   $\rho =$1, 100, and 1000  
packets and with  $nnn$-search on the for the Webgraph and Statnet. 
Lines indicate scaling dependences in Eq.\ (\ref{h-sigma}), with slopes  $\mu =1/2$ and $\mu =1$.}
\label{fig_hsigma}
\end{figure}

The observed differences that individual nodes on each  network play in the 
transport processes are further quantified by analysis of the noise fluctuations. 
We measure the number of packets $h_i(t)$ that a node $i$ processes within a 
fixed time window of $T_{WIN} = 1000$ time steps. We consider simultaneously
(multi-channel noise analysis) a set of time-series $\{h_i(t)\}$ for all 
$i=1,2, \cdots N$ nodes and $t=1,2, \cdots 500$ successive time windows. 
We determine the standard deviation $\sigma (i)$ for each of $N$ time-series 
(i.e., for each node) and plot it against the time-averaged value 
$\langle h(i)\rangle \equiv \langle h_i(t)\rangle _t$, where the average is taken over all time 
windows for each node separately. 
The  general scaling relation \cite{mb-fnd-04}
\begin{equation}
\sigma (i) \sim  \langle h(i) \rangle ^\mu \ ,
\label{h-sigma}
\end{equation}
holds for all nodes in the network for our  constant-density traffic. 
However, we find that the scaling exponents may depend on the traffic 
density in the inhomogeneous Webgraph. The results are shown in 
Fig.\ \ref{fig_hsigma} for the two network structures and three 
different traffic densities $\rho$. In particular, when the packet density is high, the number of packets 
processed by the hub nodes increases, resulting in the increased fluctuations
at these nodes. When density is comparable with the maximum buffer size 
(in our case $\rho = H =1000$), temporary congestion may occur at the main hubs and
at other nodes of large connectivity, whereas the rest of the network functions
in the free flow regime. In this situation, the noise fluctuations at more 
important nodes appear to be in the $\mu =1$ class, as opposed to the rest 
of the network, where the fluctuations follow the law with $\mu =1/2$, as  
shown in Fig.\ \ref{fig_hsigma}. Due to the absence of hubs in the Statnet 
the fluctuations remain homogeneous and in the $\mu=1/2$ class for all packet densities.

Our results demonstrate that inhomogeneous noise fluctuations may occur in the self-regulated (constant density) traffic 
away from the two universality classes defined
by \cite{mb-fnd-04}. (See also
\cite{valverde04} for traffic self-regulation with a local feedback).
In our study the explanation of these non-universal noise fluctuations is found
in how the network structure handles high density traffic.
These findings are in agreement with a recent detailed study of the empirical data
of the Internet traffic and in a specifically designed model by 
\cite{da-usftcn-06}, which revealed that when the traffic density is high
``enough'', then other parameters of the dynamics and structure play a role in
determining the noise properties.
In particular, the exponent
values may vary between 0.5 and 1, depending critically on the width of the time
window  and on the queuing times of packets
\cite{da-usftcn-06}. (Similar time-window effects are discovered  by \cite{ek-sttcsdftvs-06} in the stock market dynamics). 
Compared with earlier models which tackle the scaling of
traffic noise in the Internet
\cite{sv-itptmit-01,mb-fnd-04,valverde04,da-usftcn-06}, the  traffic model that
we study here is more complex in
that packets are navigated to specified destinations,
and that both travel times and queuing times of packets are self-consistently
determined
and appear to have power-law distributions (cf. Figs.\ \ref{fig_ptt} and
\ref{fig_ptw}). In the future we would hope to determine how 
these traffic properties contribute to the observed non-universality in noise
fluctuations.

\section{ Traffic Jamming \& Structure \label{sect_jamming}}

 A crucial feature of transport networks is that they cease to function 
efficiently when jamming of nodes occurs.  
Jamming is characterized by a drastic decrease of efficiency, in particular 
a non-stationary increase of the load as a function of $R$. Jamming occurs as a 
transition at a critical rate $R_c$.
It is of relevance to find alternative indicators or signatures, 
which are able to signal the occurrence of jamming. In particular we 
focus on exploring the  possibility to predict the 'distance' to the jamming 
transition from the activity time-series of individual nodes. 
\begin{figure}[htb]
\begin{tabular}{c}
\hskip 39pt 
{\psfig{file=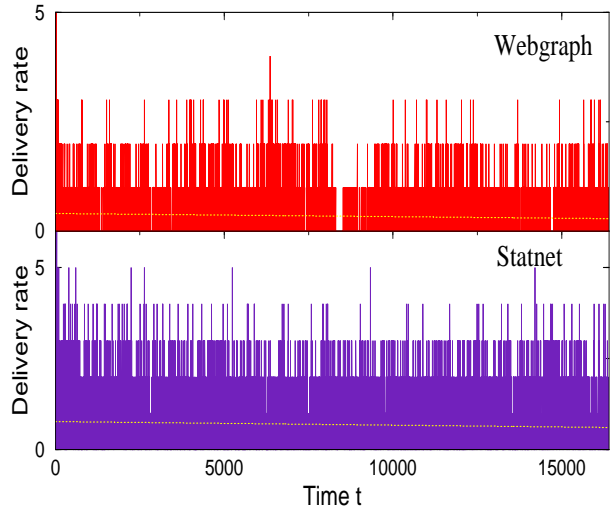,height=7cm,width=8cm}}
\end{tabular}
\caption{  Number of packets delivered per time step $n_d(t)$ against time 
$t$ for fixed
  density of moving packets $\rho =100$  in Webgraph and Statnet. Dotted lines indicate values of  average output rate $\lambda$.}
\label{fig_output-t}
\end{figure} 

As will be discussed in Section \ref{sect_optimization}, networks of different structures
perform differently when the traffic density is varied. In the constant-density traffic $\rho =100$  we show here  that the network output,
defined as the number of packets delivered per time step $n_d(t)$, 
is larger in the homogeneous Statnet compared with the scale-free Webgraph. 
In Fig.\ \ref{fig_output-t} we display the 
time variation  of the network output $n_d(t)$ rate for the two networks. 
The {\it average output rate} $\lambda $ is defined as
\begin{equation}
\lambda  \equiv <dn_d(t)/dt> \quad .
\label{def_mu}
\end{equation}
In the Webgraph $\lambda = 0.37$ we find packets per time step, 
compared with Statnet, where  $\lambda = 0.71$.  
This difference suggests that 
with an imposed traffic density and a constant number of moving packets of $\rho =100$,  
the traffic on the Webgraph topology is closer to jamming than the Statnet topology. In this Section we will explore in more detail 
the statistical signature of traffic near jamming.

\subsection{Queuing and Jamming  at Different Topologies}

The approach to the jamming regime is most appropriately studied with a {\it constant posting rate} $R$. Then the temporal variations in the network load $N_p(t)$ are given by
\begin{equation}
N_p(t) = Rt - n_d(t) \ .
\label{Np_t}
\end{equation} 
When the posting rate $R$ increases so that it exceeds the average output rate,
$R>\lambda$, the excess packets accumulate on the network leading to a systematic increase in the network load  $N_p(t)$, although the network continues to deliver packets in some way. The average accumulation rate is then obtained from 
Eq.\ (\ref{Np_t}) as
\begin{equation}
J \equiv <dN_p(t)/dt> =  R - \lambda \quad, 
\label{def_jammingrate}
\end{equation}
which is reminiscent of the congestion condition 
$J/R = 1- \lambda /R$, often found in queuing theory for a single-server queue
\cite{allen90,whitt01}. It should be stressed, however, that here the load 
$N_p(t)$ applies to the whole network, which consists of many interacting queues.
The excess load increases the queues at different nodes, depending on each node's topological centrality and its importance in the particular transport
process. Therefore,   subtle interactions between different queues and their dynamical correlations contribute to the jamming process, which are different for different network topologies.
This aspect of network congestion will be discussed in detail 
later. We first demonstrate how the statistical properties of the traffic 
change with increased posting rate by simulating traffic on the Webgraph.

\begin{figure}[htb]
\begin{tabular}{c} 
\hskip 20 pt
\resizebox{12cm}{!}{\includegraphics{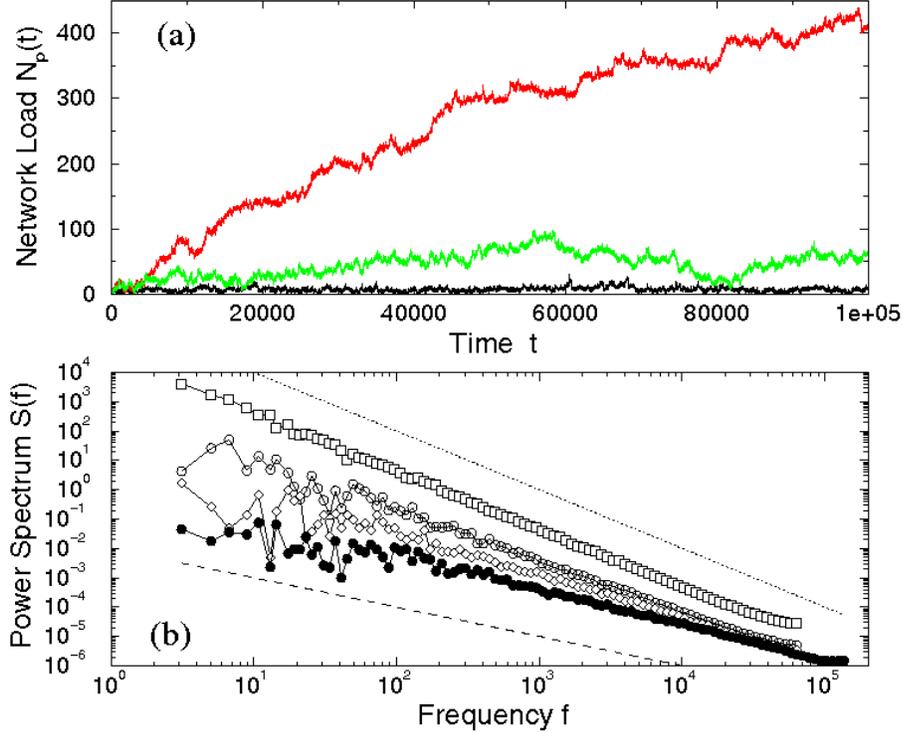}}
\end{tabular}
\caption{  (a) Network-load $N_p(t)$ time-series for transport on the Webgraph with $nnn-$search and three  posting rates indicating (bottom to top) free flow, flow with ``crisis'', and occurrence of jamming. (b)
Power-spectra of $N_p(t)$ time-series for several posting rates below jamming threshold. Dashed and dotted lines indicate limiting slopes with $\phi =1$
 and $\phi =2$, respectively. The jamming transition corresponds to $\phi =2$.}
\label{fig_wgload_t}
\end{figure}

In Fig.\ \ref{fig_wgload_t}(a) we show the time-series of the network load
$N_p(t)$ for three representative posting rates $R=$0.25, 0.35, and 0.45  on the Webgraph. Compared to the time-series of the number of active nodes in Fig.\ \ref{fig_psact}, here we take into account that each active node has a queue of packets of a length $Q_i(t) \geq 1$, which contributes to the overall network load,
$N_p(t) = \sum _i^NQ_i(t)$. 
Two of the time-series in Fig.\ \ref{fig_wgload_t}(a) are stationary, 
however, 
the average load is increasing with the posting rate $R$. Another qualitative difference is that, for the larger posting rate, $R=0.35$ in this case, 
a temporary jamming may occur that lasts for approximately $60000$ time steps, and eventually resolved by the system itself. This 'crisis'-like behavior
is one of the manifestations of the approach to  a congested flow regime
at a critical posting rate $R_c$. A systematic analysis of the traffic on 
Statnet for different posting rates $R$ reveals that the average load $<N_p(t)>$ is much larger, compared to the Webgraph, but also that the critical rate
where the jamming occurs is twice the critical rate of the Webgraph (see below). In Fig.\ \ref{fig_Np_t} network-load time-series are shown for two representative posting rates just below the respective jamming point in both networks.

\begin{figure}[htb]
\begin{center}
\begin{tabular}{c} 
{\psfig{file=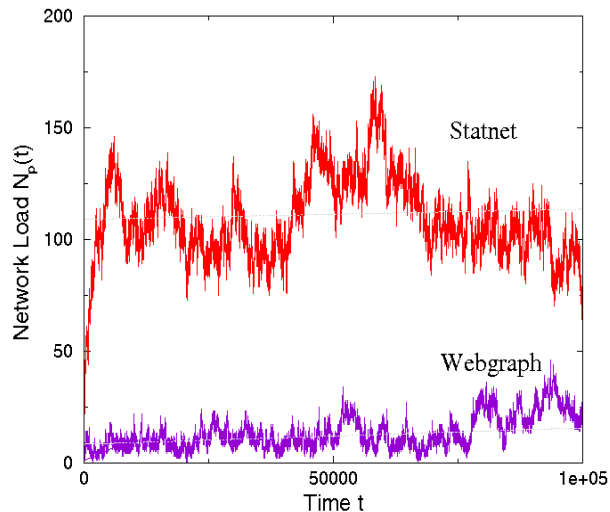,height=7cm,width=8cm}}
\end{tabular}
\end{center}
\caption{Network-load time-series close to jamming transitions in  Webgraph and  Statnet.}
\label{fig_Np_t}
\end{figure}

 As shown in Ref.\ \cite{ttr-tcntugspmdf-04}, on approaching the jamming 
point, numerous manifestations of developing congestion can be measured statistically. In particular, increased waiting times of packets and characteristic
changes in the distribution of waiting times and network loads are detected, 
as well as in the correlations of the activity and load time-series.
For the purpose of this work, we discuss  only the systematic changes in the
network-load time-series.  The power-spectra of the network-load time-series for the traffic on Webgraph are shown in Fig.\ \ref{fig_wgload_t}(b) for different values of the posting rate in the range $R=0.005$ to $R=0.4$. The remarkable feature of the spectra is the systematic decrease in the correlations (antipersistency), measured by the increased scaling exponnet $\phi$ in Eq.\ (\ref{def_ps})
from $\phi =1.18$ at lowest considered posting rate, to $\phi =2$, at the
jamming threshold. This gives the numerical evidence that the onset of congested state
is characterized by a loss of long-range correlations in packet streams.
This feature seems to apply generally. The transition to the congested
state and the properties of the traffic in the congested regime, however, 
depend on the topological details of the network. 

In order to avoid possible confusion regarding the role of the transition
point to the congested state, here we would like to stress that the network-load
time series in our model exhibits long-rage correlations (with an exponent $\phi
< 2$) in
 {\it a wide range of posting rates below the jamming transition}. As we show in
the next subsection, these correlations are strictly related to the structural
complexity of the network.

\subsection{Transition to Congested Traffic State}
\begin{figure}[htb]
\begin{tabular}{c}
\hskip 36pt 
{\psfig{file=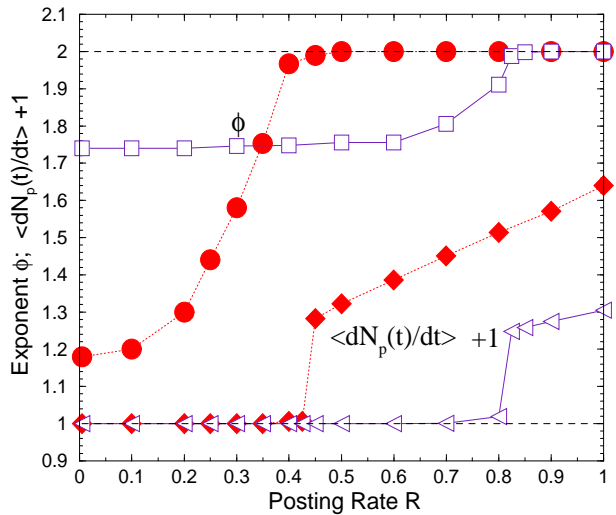,height=7cm,width=8cm}}
\end{tabular}
\caption{(upper curves) Correlation exponent $\phi$  defined in 
Eq.\ (\ref{def_ps})
and (lower curves) jamming rate $<dN_p(t)/dt>$ (shifted by 1) for different 
values of  posting rates $R$  in
traffic on the  Webgraph (filled symbols) and Statnet (open symbols).}
\label{fig_phi-R}
\end{figure}

Technically, the transition point can be identified
with $\phi (R_c) \to 2$ for a given network geometry. A characteristic slow-down
of the dynamics at a congested node, typically a hub in the Webgraph, occurs 
in that
only one packet can move-in from one of the neighbour nodes, and a large number of packets at other neighbouring nodes are waiting until one packet 
moves-out the hub in the next time step. 
Study of long time-series for different posting rates reveals how the de-correlation occurs on different network structures. In Fig.\ \ref{fig_phi-R} we 
show
values of the exponent $\phi$ against $R$ in the case of Webgraph and Statnet.
It shows first that large variations in the strength of correlation occur in 
load time-series on the Webgraph, in contrast to the time-series on the Statnet, which exhibit weaker but stable correlations over a wider range of posting rates.  
The values of the critical rates are $R_c \approx 0.4$ 
for traffic on the Webgraph, whereas $R_c\approx 0.8$ for the Statnet,  
roughly coincide  with the time-series decorrelation rates.
 
Additional quantitative characteristics of the transition to the congested state can be obtained by measuring the average jamming rate, $J$, which is
 defined in Eq.\ (\ref{def_jammingrate}). In Fig.~\ref{fig_phi-R} we show systematic dependences of the jamming rate $J$ from the externally imposed
posting rate $R$ for traffic on both types of networks. It reveals that the critical posting rates $R_c = 0.4375 \pm 0.0125$ for traffic on
the Webgraph, and $R_c=0.75 \pm 0.0125$ in the case of Statnet, at which the jamming transition starts to appear are closely associated with a complete decorrelation in the respective time-series ($\phi $ reaches the value of 2 in both cases).
Apart from the differences in values of the critical rates $R_c$, the 
onset of jamming seems to occur abruptly in both networks. Additional differences due to network topology are seen in the character of the jammed traffic. Namely, the slopes of the curves for $J(R)$ above the jamming point in each case are different. This suggests that 
in the congested state the delivery rate $\lambda$, according to Eq.\ (\ref{def_jammingrate}) drops to $\lambda /R = 1 - J/R$, where $J/R=0.7$ for the case of Webgraph, and $J/R = 0.5$ for the case of Statnet, are slopes of the curves $J(R)$ in Fig.~\ref{fig_phi-R}. Therefore, in the congested state the Webgraph continues to process about 30\% of posted packets, whereas the Statnet 
manages to deliver about 50\%.

At this point it is interesting to compare the constant density traffic, $\rho =100$ studied in previous section, with the picture of jamming on the two topologies.
In the constant density traffic an effective posting rate $<R>$ emerges,  which keeps the balance of the network's output rate.  According to 
Fig.~\ref{fig_output-t}, for $\rho = 100$ the effective posting rates,  $<R> = \lambda =0.37$ for the Webgraph,
and  $<R> = \lambda = 0.71$ in the case of Statnet, are below the respective
jamming point in both topologies.

\section{Optimized Transport \label{sect_optimization}}
Perhaps the most important question in transport optimisation is to identify the  optimal performance of a network, 
once certain constraints such as search mode and queue protocols are specified. 
Here we review two ideas that address this question from different standpoints. 
The first develops a method to identify optimal paths in different networks 
by using maximum flow trees \cite{tt-tnmfstgsn-06}, the other actually optimises network structure 
for a given protocol and search depth \cite{gdvca-ontlsc-02}.

 \subsection{Optimal Paths on Fixed Topologies }

\begin{figure}[thb]
\begin{center}
\begin{tabular}{cc} 
{\psfig{file=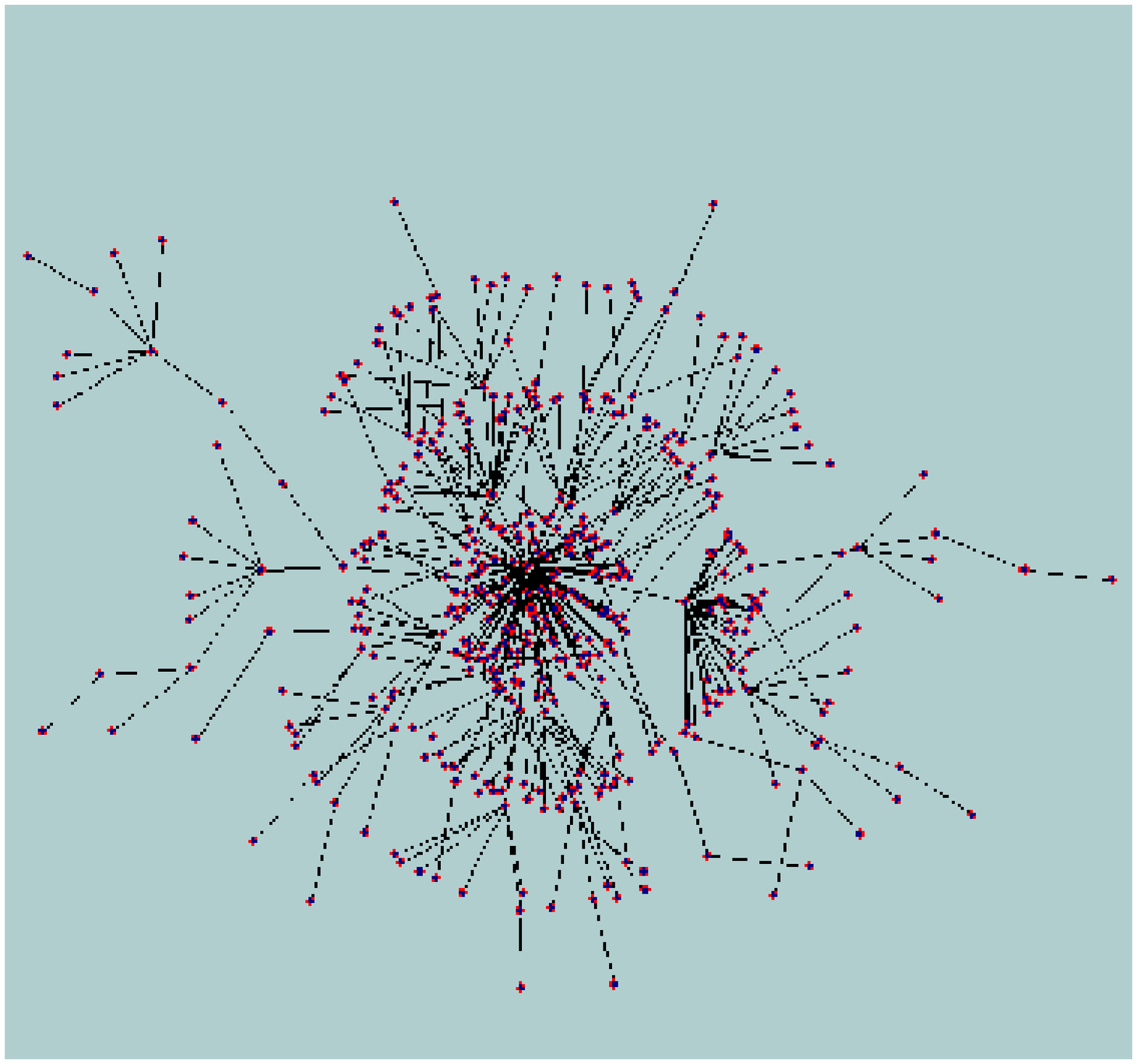,height=6.4cm,width=6.4cm}}&
{\psfig{file=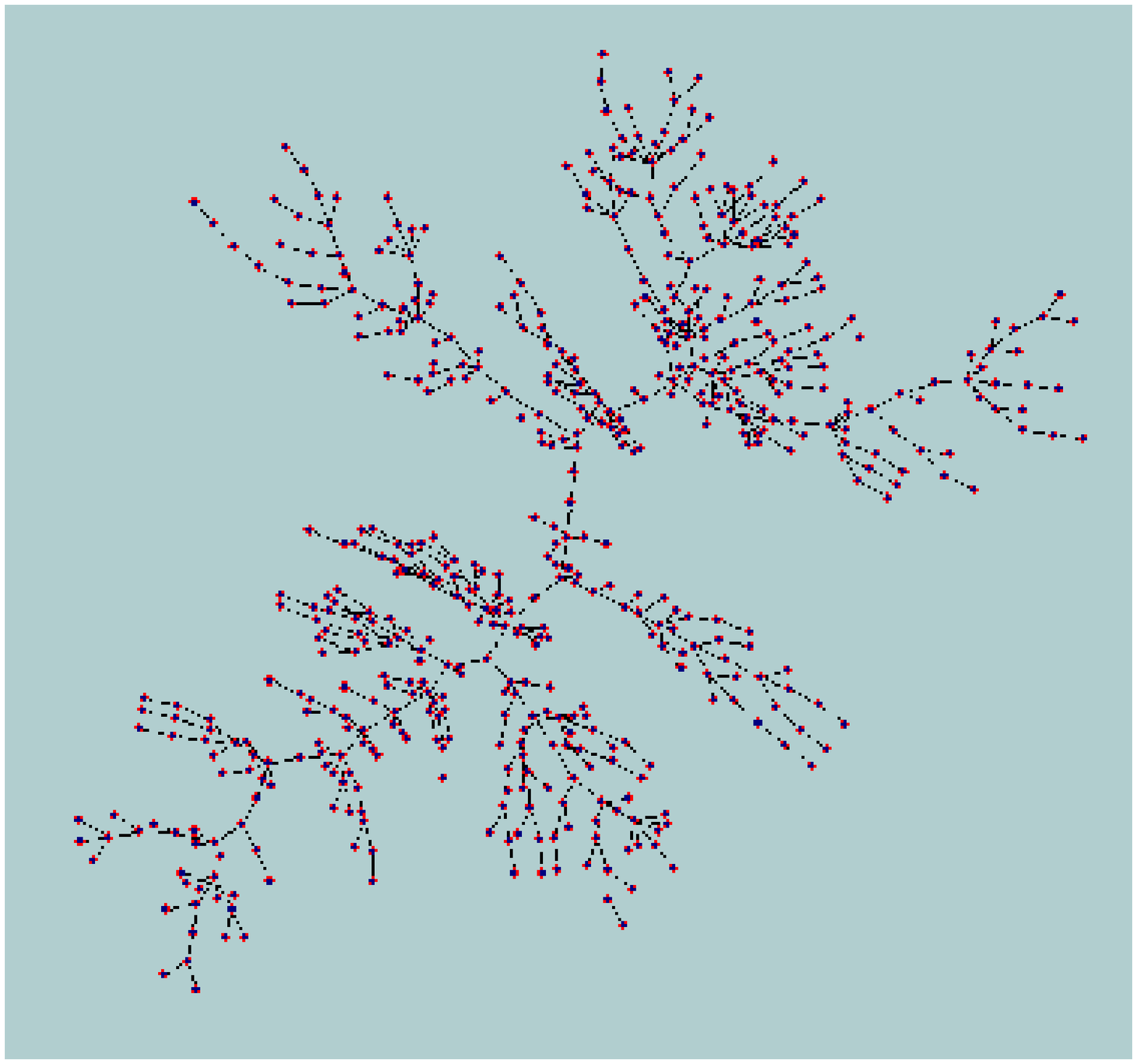,height=6.4cm,width=6.4cm}}\\
{\Large $(a)$} &{\Large $(b)$}
\end{tabular}
\end{center}
\caption{ Dynamic  maximum-flow spanning tree   traffic density $\rho =1$
on the cyclic scale-free Webgraph (a) and  homogeneous Statnet (b).}
\label{figgraph_mst}
\end{figure}

By simulating a large number of packets we record the number of walks 
along each link (dynamic flow) and through each node (dynamic noise) in the network.
Obviously, the inhomogeneity of nodes with respect to their local network environment, as shown in Fig.\ \ref{fig_conn-centrality},  
makes the flow and noise fluctuations different on each network structure. 
To obtain a quantitative 
analysis of flow  on the network links, here we construct a {\it maximum-flow spanning tree}, 
on which each node is connected to the rest of the network nodes via its 
maximum-flow link. Implementing a greedy algorithm, we determine the trees 
from maximum-flow in the $\rho =1$ limit on the  two networks. 
The trees are shown in Fig.\ \ref{figgraph_mst}.

The structure of these trees reflects both the underlying network 
geometry and how that geometry effects transport with given navigation 
rules--local $nnn$-search. In the case of the Webgraph the tree exhibits  
a scale-free topology, suggesting a certain degree of
compatibility between the traffic and the structure. Similarly, for the 
Statnet 
the tree shows some degree of inhomogeneity that corresponds to the weaker inhomogeneity of the underlying graph.

\begin{figure}[htb]
\begin{center}
\begin{tabular}{c} 
{\psfig{file=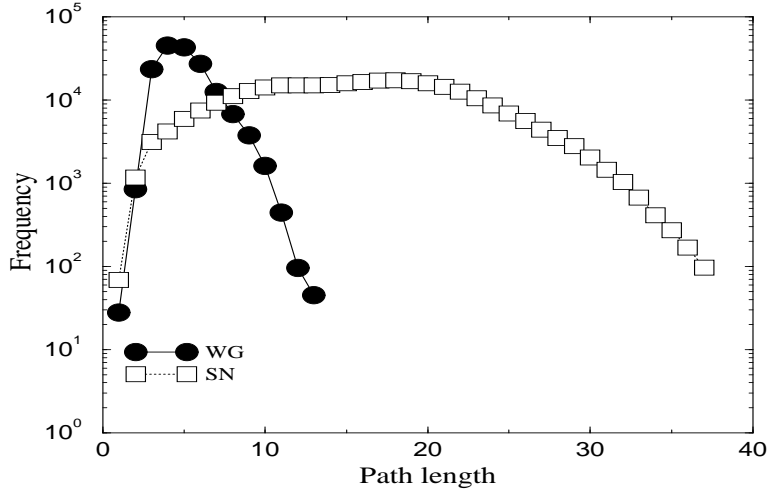,height=2.6in,width=4.0in}}
\end{tabular}
\end{center}
\caption{Distance distributions on the maximum-flow spanning trees for the Webgraph and the 
Statnet for $\rho =1$, and $nnn$-search [Tadic and Thurner, 2006].}
\label{fig_pdmst}
\end{figure}
The maximum-flow spanning trees represent the union of 
maximum-flow paths on the underlying network structure. 
In Fig.\ \ref{fig_pdmst} we show distribution of the lengths of all such paths 
on the two trees that are shown in Fig.\ \ref{figgraph_mst}.
Once again, differences in the graph topologies and thus in 
their maximum traffic trees manifest themselves in the statistics of the maximum-flow paths. For instance, 
the average distance along such paths on the Webgraph and Statnet differs by a factor of about 
5, the maximum distance differs by a factor of about 3.
 
The message is that for a fixed navigation protocol the underlying network structure 
determines the topology of {\it optimal paths} for packet transport. These paths
consist of links that appeared to be {\it locally optimal} choices for packets,
without any global feedback. 
The quantitative differences in these 
topologies can be determined in the limit of non-interacting packets (as shown in Fig.\ \ref{figgraph_mst}). However, how these optimal paths
 are used by packets when the traffic density is increased, again depends on
the properties of the network queues. 
In particular,  the traffic efficiency due to
very short paths between pairs of nodes, as on scale-free topologies, can be 
hindered by the occurrence of  large queues at hubs, discussed in Sect.\ \ref{sect_jamming}. In the following  we show what  
 network topologies emerge as {\it globally optimal structures}  in 
relation to varying traffic density \cite{gdvca-ontlsc-02}.

\subsection{Traffic Optimization by Network Restructuring}

In Sects.\ \ref{sect_flow_simulations} and \ref{sect_jamming} we have demonstrated that  transport efficiency  on a given network topology may crucially depend on the traffic density. Specifically in strongly inhomogeneous networks, such as the Webgraph, the structural characteristics which are 
advantageous at low traffic density---occurrence of powerful central nodes, may appear as weaknesses at high density traffic. 
In this subsection we would like to draw attention to a more systematic way to obtain an optimal structure for a given density using network reconstruction. 
The suitable formalism was introduced in Ref.\ 
\cite{gdvca-ontlsc-02}, where global structure optimization is performed starting from data about individual  packet flow. In this respect, 
the formalism 
simultaneously accounts for both search and congestion aspects of the information flow on networks. The basic arguments 
 are presented below. 
For more details see the original Ref.\ \cite{gdvca-ontlsc-02}.

  As before the focus is 
on a single information packet at node $i$ whose destination is node $k$.
The probability for the packet to go from $i$ to a new node $j$ in its next movement is $p_{ij}^k$.
In particular,
$p_{kj}^k=0$,  $\forall j$ so that the packet is  removed as soon
as it arrives at its destination. 
The probability $p_{ij}^k$  depends on the 
network topology, given by the elements of the adjacency matrix $c_{ij}$, 
and on the search algorithm. 
The three cases with local search modify the transition probability $p_{ij}^k$ in the following way: 
For a random walk (searched depth zero) we have 
\begin{equation}
p_{ij}^k=\frac{c_{ij}}{\sum_jc_{ij}} \quad , 
\label{pijk_RD}
\end{equation}
and for a random walk with nearest neighbour search 
\begin{equation}
p_{ij}^k=c_{ik}\delta_{jk}+(1-c_{ik})\frac{c_{ij}}{\sum_jc_{ij}} \quad . 
\label{pijk_nn1}
\end{equation}
Finally, for the the  $nnn$-search used in Sects.\ \ref{sect_flow_simulations} and \ref{sect_jamming}, it  corresponds to
\begin{equation}
p_{ij}^k=\left\{
\begin{array}{ll}
 \delta_{jk} & \mbox{if $c_{ik}=1$}\\
 \frac{c_{ij}c_{jk}}{\sum_jc_{ij}c_{jk}} & \mbox{if $c_{ik}=0$ and 
$\sum_jc_{ij}c_{jk}>0$}\\
 \frac{c_{ij}}{\sum_jc_{ij}} & \mbox{if $c_{ik}=0$ and 
$\sum_jc_{ij}c_{jk}=0$}\\
\end{array}
\right .
\end{equation}

Following the Ref.\ \cite{gdvca-ontlsc-02}, one can then determine the dynamic betweenness of nodes and, subsequently, the network load $N_p(t)$, under certain fairly general conditions. 
When the search is Markovian, i.e. $p_{ij}^k$
does not depend on previous positions of the packet, the
probability of going from $i$ to $j$ in $n$ steps is given by
\begin{equation}
P_{ij}^k(n)=\sum_{l_1,l_2,\dots,l_{n-1}}p_{il_1}^k p_{l_1l_2}^k\cdots p_{l_{n-1}j}^k \quad .
\end{equation} 
This definition allows calculation of the average number of times, $b_{ij}^k$,
that a packet generated at $i$ and with destination at
$k$ passes through $j$. This can be expressed in matrix notation as
\begin{equation}
b^k=\sum_{n=1}^\infty P^k(n)=\sum_{n=1}^\infty
\left(p^k\right)^n=(I-p^k)^{-1}p^k  \quad  , 
\label{bij}
\end{equation}
where the matrices $b^k$, $P^k(n)$ and $p^k$ have elements $b^k_{ij}$, $P^k_{ij}(n)$ and 
$p^k_{ij}$ respectively and $I$ is the identity matrix.
The effective (dynamic) betweenness, $B_j$, of node $j$  is then 
defined as the
sum over all possible origins and destinations of the packets on the graph,
\begin{equation}
B_j=\sum_{i,k}b_{ij}^k \quad .
\label{B}
\end{equation}
When the search
algorithm is able to find the minimum paths between the origin and destination nodes, the
{\it effective} betweenness will coincide with the {\it topological}
betweenness \cite{n-scnspwnc-01,zlw-cscn-05,Latora04}, considered in Sec.\ \ref{sect_flow_simulations}.

If packets are generated at random and independently with
a probability $R$, or at each node with probability $r=R/N$, 
and if the queuing discipline is given by a random queue M/M/1, rather than the more complicated 
LIFO used in Section\ \ref{sect_flow_simulations}, it can be shown  that the
time averaged load of the network is \cite{allen90,gdvca-ontlsc-02}
\begin{equation}
\overline{N_p}=\sum_{j=1}^N\frac{\frac{r B_j}{N-1}}{1-\frac{r B_j}{N-1}} \quad.
\label{load}
\end{equation}
This solution has two interesting limiting cases: For small values of
$r$ 
the average load is  be proportional to the average effective distance
\cite{gdvca-ontlsc-02}.
On the other hand, when $r$ approaches a critical rate $r_c$, most of the load of
the network comes from the most congested node, and therefore
\begin{equation}
\overline{N_p}\approx\frac{1}{1-\frac{r B^*}{N-1}} \ ;\quad\quad r\rightarrow r_c \quad ,
\label{N2}
\end{equation}
where $B^*$ is the effective betweenness of the most central
node, and  
assuming  
that the jump probabilities $p_{ij}^k$ do not depend on the congestion
state of the network.

\begin{figure}[thb]
\begin{center}
\begin{tabular}{cc} 
{\psfig{file=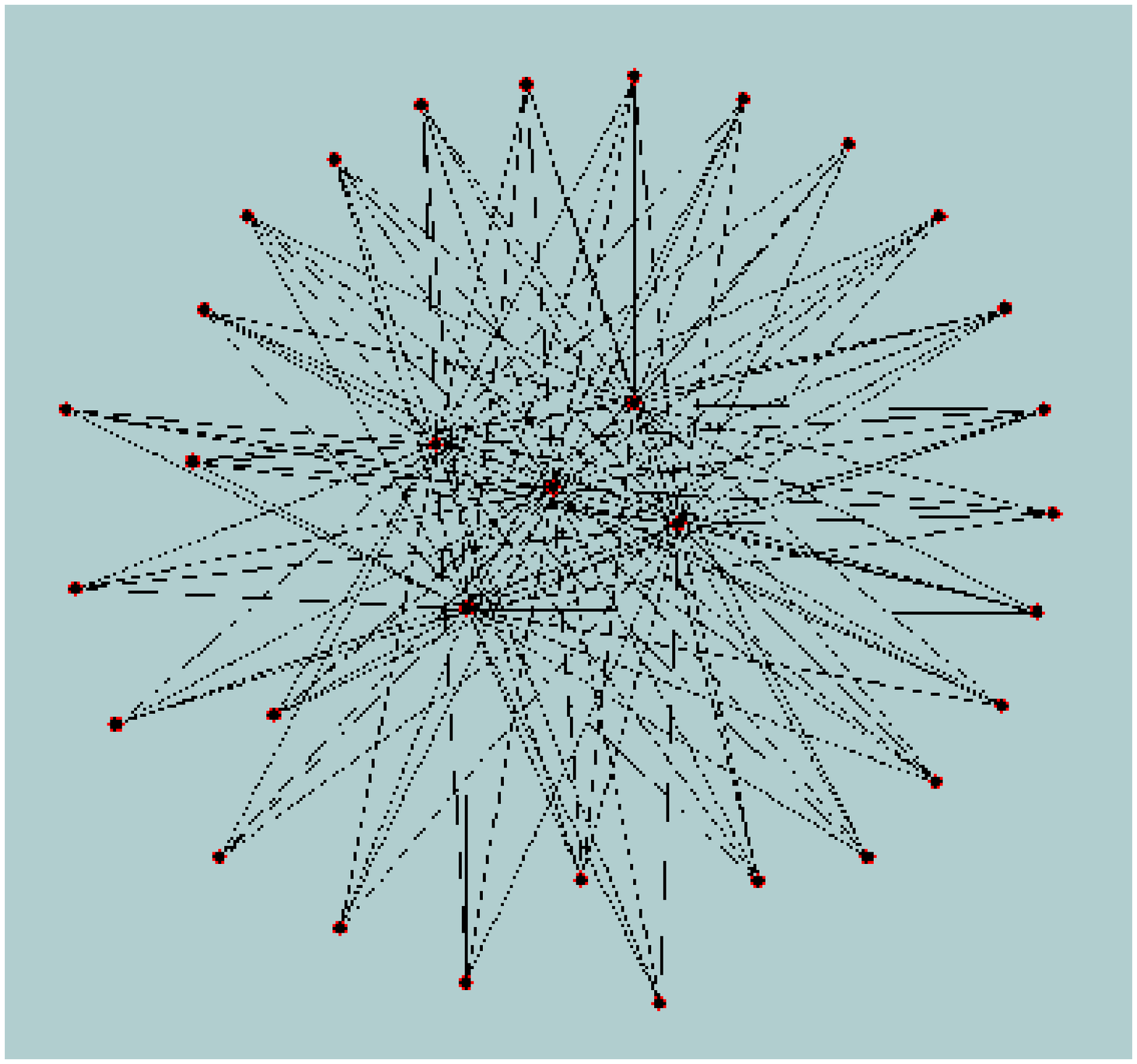,height=6.4cm,width=6.4cm}}&
{\psfig{file=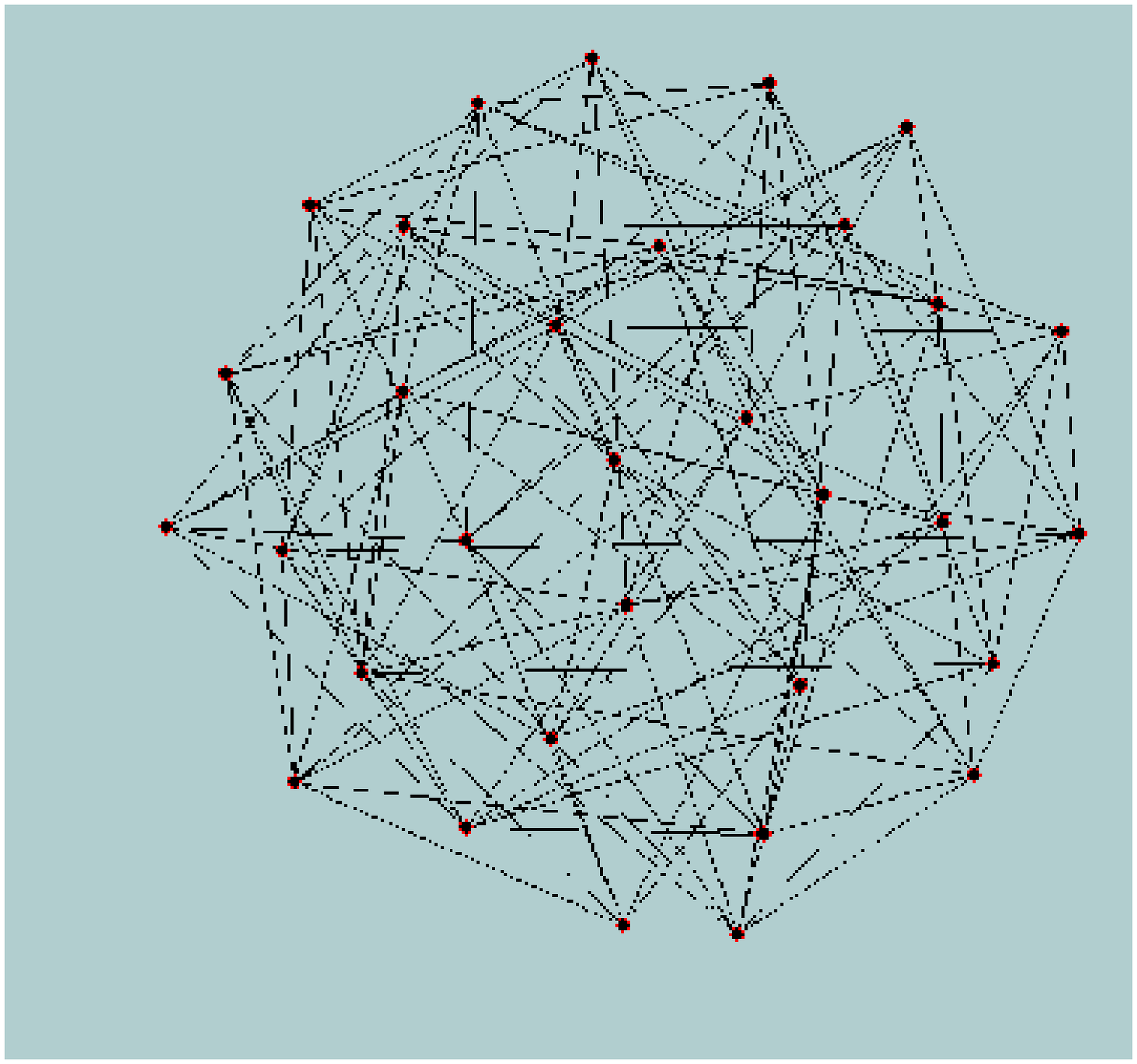,height=6.4cm,width=6.4cm}}\\
{\Large $(a)$} &{\Large $(b)$}
\end{tabular}
\end{center}
\caption{Optimal networks for low-density traffic (a)
and for high-density traffic (b). (Data: courtesy A. D\'{\i}az-Guilera.)}
\label{figgraph_opt_structures}
\end{figure}

Equation (\ref{load}) relates a dynamical variable, the load, with topological
properties of the network and  of the search algorithm.
Hence, the dynamical optimization procedure of finding the structure that 
gives the minimum load is reduced to a topological optimization procedure 
where the network structure is characterized by its effective (dynamic)
betweenness 
distribution. In \cite{gdvca-ontlsc-02} the problem of finding 
optimal structures was considered for a purely local search, Eq.\ (\ref{pijk_RD}), using 
a generalized simulated annealing procedure, as described in 
\cite{tsallis94,penna95}. 
Surprisingly, it was found that there are only two types
of structures that can be optimal for a local search process:
star-like networks for posting rates below a characteristic rate $R^*$,
$r<r^*$ and homogeneous networks for
$r>r^*$. The networks are shown in Fig.\ \ref{figgraph_opt_structures}.

The network structures which result from the optimization process for two 
limits of the traffic density, Fig.\ \ref{figgraph_opt_structures},  exhibit
a remarkable similarity with graphs considered in Sec.\ \ref{sect_flow_simulations}, in particular,
with their {\it core-graphs}  (graphs without single-link nodes). This 
comparison helps us to identify the main topological property of the 
graphs in 
Fig.\ \ref{figgraph} that is responsible for the transport  efficiency. 
In particular, it 
suggests that the large clustering in the Webgraph near the main hubs
is responsible for the increased efficiency of transport at low traffic 
density \cite{tt-isdsn-04,ttr-tcntugspmdf-04} (see also Section \ref{sect_jamming})
compared to other scale-free networks. 
Similarly, the absence of clustering and a narrow distribution of the  
connectivity in the Statnet, makes it suitable to support a large-density 
traffic, in analogy to the homogeneous network in Fig.\ \ref{figgraph_opt_structures}.

\section{ Open Problems and Conclusions \label{sect_openproblems}}

\subsection{Summary}
We have shown using two complementary approaches that two different classes of 
network structure give 
rise to different traffic properties. The classes of network structures 
we considered were 
scale free graphs with a significant hub node and high clustering (called Webgraph), and 
a much more homogeneous network with a quickly decaying degree distribution 
(called Statnet). 
These classes were identified in two ways. 
Firstly, by directly implementing a fairly realistic protocol for packet transport on these networks we demonstrated with a number of quantitative characteristics that the traffic behaves differently on the two networks.
Secondly, making use of the results of network optimization in
 \cite{gdvca-ontlsc-02}, we concluded  that these two network classes appear to be representative of optimal  
structures for low and high traffic density, respectively. 

A comparative study of packet transport on these two networks revealed  
 how traffic properties depend on the network structure and on the packet density.
We considered both free moving and congested traffic, and driving conditions with both
a constant packet density and with a constant posting rate. 
A summary of the traffic characteristics identified is given below:

\noindent
{\it Power-law tails:} In the statistics of individual packets
the distributions of  travel time, $P(T)$, and waiting time, $P(t_w)$,
exhibit  power-law tails on both network types, however, with different slopes (cf. Figs.\ \ref{fig_ptt}, \ref{fig_ptw}). Further
 differences appear at short travel times, where transport on the Webgraph mostly follows the shortest path between source and destination nodes, while the
times up to a characteristic time $T_0 \approx 10$ steps are all equally probable on Statnet. Consequently, the distribution $P(T)$ fits
a $q-$exponential \cite{tsallis88} in the homogeneous Statnet and a true power-law
in the scale-free Webgraph. The distributions of return-time, $P(T_r)$,
 has  power-law tails with a large slope on both networks. The main differences between the networks appear at short return times (see Fig.\ \ref{fig_prett}), where the inhomogeneous betweenness of nodes in the 
Webgraph play an important role.

\noindent
{\it Universal noise fluctuations:} The traffic noise averaged over a suitably chosen time window follows the universal law in Eq.\ (\ref{h-sigma}) in both network classes.
 In the free flow regime we find $\mu =1/2$ law in both network types, 
regardless of the differences in their  homogeneity and 
clustering characteristics.
However, when the traffic jamming occurs at large density, the law changes towards $\mu =1$,  for the busiest hub nodes in the inhomogeneous scale-free Webgraph, whereas no qualitative changes are observed in the homogeneous Statnet. 

\noindent
{\it Antipersistent time-series:} The network load time-series $N_p(t)$
in both network types are found to be antipersistent in the free flow regime.
The degree of correlations, however, are considerably different
 (Fig.\ \ref{fig_phi-R}), depending on
traffic density (or posting rate). The noise correlations on  Webgraph 
vary from very strong, at low density,  to weak correlations, at the jamming threshold 
$R_c \approx 0.4$. In the case of homogeneous Statnet the noise correlations
are generally weaker but stable in a wide range of posting rates until the jamming threshold, $R_c \approx 0.8$, when only short-range correlations 
survive.

\noindent
{\it Jamming transition:} A jammed traffic regime, characterized by 
unbalanced increase of network load 
occurs on both networks at their critical  posting rates $R_c$, mentioned above, Fig.\ \ref{fig_phi-R}.  
The appearance of the nonzero jamming rate $\lambda$ (Eq.\ \ref{def_mu})
is quite sudden for 
finite sized networks of both classes. For a given network type 
the value of the critical  threshold $R_c$
 depends strongly on the efficiency of the search algorithm \cite{tt-isdsn-04}. Specifically, 
the critical rate $R_c \approx 0.4$ in   traffic with
 $nnn$-search on the Webgraph is particularly high compared to random diffusion on the same graph  or traffic on scale-free trees, where the critical jamming rate is of the order $R_c\approx 10^{-3}$  \cite{tt-isdsn-04}.

\noindent
{\it Traffic in the congested state:}  The advantage of the homogeneous structure 
at high traffic density continues to be seen in the congested regime. However, the difference in the efficiency is only about  20\%. This can again be attributed to the efficient $nnn$-search of the Webgraph geometry.

\noindent
{\it Optimal flow paths:} The union of the optimal paths learnt from the traffic history, summarized at the maximum-flow spanning trees of the two network types in Fig.\ \ref{figgraph_mst} shows the advantage of the scale-free structure  at low density traffic. Due to the efficient search, the maximum-flow tree retains the
scale-free character and the statistics of the optimal path lengths in Fig.\ \ref{fig_pdmst} resemble those of the topologically shortest paths on the 
graph.
On the other hand, paths are much longer on the homogeneous Statnet, 
independent of the traffic density.

For a fixed network geometry further, although limited, optimization  
may be achieved  by improving the search algorithms \cite{krt-libaptcn-06}.
Transport processes on other types of networks, cellular networks \cite{st-tphpgsfl-06}, 
gradient networks \cite{tb-jlsfs-04,pl-jcgn-05},
river networks \cite{banavar99}, or generally trees and directed graphs, are subject 
of additional constraints that may result in qualitatively different behavior.
These networks are not considered in the current work.

\subsection{A comment regarding 'realistic' protocols}
Among various transport networks the Internet is an particularly
accessible and 
  attractive example. In recent years measurements in both 
the Internet structure 
and the information packet TCP/IP traffic dynamics on the Internet have 
been carried out  and their inter-relations discussed 
\cite{c-ofncnt-94,tts-cbofnit-96,ttf-dptoitf-00,wgjps-spicec-02,abe-suz-03,dallrstw-ryfni-05}.
These have either considered ping time 
statistics, that is a packet's round trip time from source to 
destination and back again, or the statistics of the load on a 
particular server, router or cable.
Almost all these studies have found scaling laws and 
long range correlations in a range of traffic time-series 
\cite{c-ofncnt-94,tts-cbofnit-96,ttf-dptoitf-00}.  
Examination of the power spectra has allowed the identification of 
two different regimes with free flow or jammed traffic.
It was shown \cite{wtsw-IEEE-97} that the power-law behaviour of the 
distribution of packet inter-arrival times has a significant impact on packet 
queue statistics, and consequently on the overall traffic performance.

In addition, the  scale-free nature of the Internet structure  
\cite{vpsv-lstdpi-02,dallrstw-ryfni-05} has been fully characterised.  
The  structural characteristics  on the autonomous systems level, 
in particular the degree distribution, high clustering and 
link-correlations,  are 
statistically  similar to those of the prototype network Webgraph discussed 
in Sec.\ \ref{sect_flow_simulations}. 
Beyond the practical importance of research into the properties of 
the Internet, the underlying cause
of self-similarity and criticality in the Internet's structure and information traffic is still a subject of debate involving researchers in a broad range of scientific disciplines. Among the goals of this research is to determine universality classes of the scale-free behaviour and to unravel the mechanisms of the 
structure--dynamics interdependences.

Examining our results in Sec.\ \ref{sect_flow_simulations}, 
obtained with the  local $nnn$-navigation algorithm, we can conclude that  
the statistical features of the 
traffic, in particular the power-law distributions, the degree of 
correlations in  the packet streams,
and the onset of jamming, are  related to the actual Internet 
structure. Furthermore, our results suggest that no substantial improvement in traffic efficiency will be achieved by implementing a long-range search beyond
the ``critical horizon'' of the scale-free network, which is 
 depth level two in our Webgraph (more detailed analysis is given in 
\cite{tt-statsfn-05}). In the study by \cite{valverde04}, who first introduced the term ``critical path
horizon'' in this context, the sharp transition was found at the depth level
four.
The difference in the critical horizons in these two scale-free networks can be
attributed to the better "searchability" of the correlated scale-free Webgraph,
compared to uncorrelated scale-free network used in \cite{valverde04}.

Of crucial importance is the centrality of nodes, which pre-determines the size of queues, and hence the traffic jamming and an increased risk of packet loss. 
The onset of traffic congestion seems to occur suddenly in the prototype
network, and is probably related to the size of the giant cluster. 
However, we observed a kind of traffic ``crisis'' behaviour with large load fluctuations before the actual jamming starts. 
The statistical indicators of the traffic behaviour, determined in previous sections,
 systematically change with the increased network
load (traffic density). Hence 
the following four  phases can be clearly identified: Free flow, crisis, jamming threshold and congested traffic. Therefore, an effective control mechanism, that would eventually lead to increased traffic efficiency and security, 
may be developed through a   
systematic monitoring  of one or more of these indicators.

\subsection{Some Open Theoretical Problems in Transport on Networks}

Quantitative  properties of transport on a network topology depend 
in different ways  on the {\it network geometry} and on {\it search algorithm} and {\it type of queues} on that geometry. The actual dependence on the search and queuing cannot be considered independently of the underlying network structure \cite{tt-statsfn-05}. Another subtle factor which determines the transport properties is  the {\it driving mode}. Specifically, large fluctuations of the posting rate may influence noise properties. More seriously, they may
drive the network out of a stationary flow, where some statistical properties
cannot be determined  mathematically correctly. 

For the theoretical purposes we have implemented in Sec.\ \ref{sect_flow_simulations} a non-invasive  {\it self-consistent} driving, which preserves
the number $\rho$ of moving packets  in time. The stationarity of time-series is thus guaranteed, and jamming does not occur when $\rho$ is not too big, as
shown above. In this regime we can determine numerically the correct
statistical properties
of the traffic. In the limit $\rho \to 1$ the outcomes are directly
attributed  to the network structure. The results of numerical simulations are given in Sec.\ \ref{sect_flow_simulations}. However, clear theoretical 
concepts behind these numerical results still remain elusive, in particular 
related to the following questions:
\begin{itemize}
\item Universality classes of travel time distributions.  

\item Power-laws in the waiting time distribution.

\item Queue interactions in the correlated network environment.

\item Return-time distributions structural dependences
 and occurrence of the $q-$exponential.

\item Robustness of the universal noise fluctuations,
related to traffic density and network structure.

\item Universality of the dynamic jamming transition.
\end{itemize}

In traffic models, as in many other dynamical systems, power-law behaviour is 
often attributed to (dynamic) phase transitions, separating different 
attractors of the dynamics. 
A particular example of this is the jamming transition in 
the 
traffic model on a compact lattice studied by \cite{sv-itptmit-01}. However, when 
the 
underlying network is sparse and strongly inhomogeneous, additional correlations may become relevant, leading to attractive states with 
self-similar dynamics away from the phase transition. In this paper, by comparing  
traffic on different networks, we have found substantial evidence 
that the self-similar dynamic properties 
(at and away of the jamming transition) are shaped by the network 
geometry.

In conclusion, the  dynamic characteristic of transport processes on networks,
appear to be, to a large extent, predesigned by the 
underlying network structure, where certain structural characteristics 
play a  dominant role. 
However, these structure--function interdependences are strongly
determined by the dynamic  conditions, in particular traffic  density. 
In this respect, two large classes of networks with an optimal function
are now identified---clustered scale-free networks, and homogeneous, or 
weakly structured, unclustered networks.  With the large-scale numerical 
simulations in this work we have shown that different
quantitative characteristics of the traffic emerge in these two network classes. This leaves open the question about the active principle which shapes the dynamics within  each of these network classes. 
 We hope that our investigations will initiate further theoretical and 
practical research to address this question.

\section*{Acknowledgments} 
BT acknowledges support from the Program P1-0044 of the Ministry of Higher Education, Science and Technology of the Republic of Slovenia; 
ST for FWF Project P17621, Austria.
Further, partial support from the COST Action P10 Physics of Risk, is acknowledged.

\bibliography{transport}

\end{document}